\documentclass[prd,aps,nofootinbib,floatfix,11pt]{revtex4}
\usepackage{amsmath,graphicx,epsfig,amssymb,dsfont,mathtools,}
\usepackage[usenames]{color}
\usepackage{ulem} 
\usepackage{bigstrut}
\usepackage{slashed}
\usepackage{multirow}
\usepackage{subfigure}

\allowdisplaybreaks

\begin{document}\title{Resonance contributions in {\boldmath$B^-\to K^+K^-\pi^-$}\\
  within the light-come sum rule approach}

\author{Yu-Ji Shi$^{1}$~\footnote{Email:shiyuji92@126.com},
  Ulf-G. Mei{\ss}ner~$^{1,2,3}$~\footnote{Email:meissner@hiskp.uni-bonn.de} and Zhen-Xing Zhao~$^{4}$}
\affiliation{$^1$ Helmholtz-Institut f\"ur Strahlen- und Kernphysik and Bethe Center \\ for Theoretical Physics,
  Universit\"at Bonn, 53115 Bonn, Germany\\
  $^2$ Institute for Advanced Simulation, Institut f{\"u}r Kernphysik and J\"ulich Center for Hadron Physics,
  Forschungszentrum J{\"u}lich, D-52425 J{\"u}lich, Germany\\
$^3$ Tbilisi State University, 0186 Tbilisi, Georgia\\
$^{4}$ School of Physical Science and Technology, Inner Mongolia University, Hohhot 010021, China}

\begin{abstract}
In this work, we study the resonance contributions to the decay $B^-\to K^+K^-\pi^-$, which
is dominated by the scalar $f_0(980), K_0^*(1430)$, vector $\rho(1450), \phi(1020), K^*(892)$ and
tensor $f_2(1270)$ resonances. The three-body decay is reduced to various quasi two-body decays,
where the $B$ meson firstly decays into a resonance and a pion or a kaon, subsequently the
resonance decays into the other two final-state mesons. The quasi two-body decays are calculated
within the light-cone sum rule approach utilizing the leading twist $B$ meson light-cone distribution
amplitudes. Finally, the decay branching fraction contribution from each resonance are calculated. Some of
them are compatible with experiment or previous theoretical works. Including the effects of non-resonant and
final-state interactions, we also evaluate the total branching fraction.
\end{abstract}
\maketitle

\section{Introduction}

The non-leptonic three-body decays of $B$ mesons are an ideal platform for the study of direct CP violation.
Nowadays, using the Dalitz plot analysis, the LHCb collaboration has measured  direct CP violation in
charmless three-body decays of $B$ mesons \cite{LHCb:2013ptu,LHCb:2013lcl,LHCb:2014mir}, where 
evidence for inclusively integrated CP asymmetries has been found. 
Generally, the decay amplitude is a crucial quantity to obtain the CP asymmetries. The decay amplitude
of $B^-\to K^+K^-\pi^-$ consists mainly of three components, which involve resonant and non-resonant (NR)
effects as well as final-state interaction (FSI) effects. Recently, LHCb and BaBar have studied the
decays $B^{+} \rightarrow \pi^{+} \pi^{-} \pi^{+}$, $B^{+} \rightarrow K^{+} K^{-} \pi^{+}$ and
$B^{\pm} \rightarrow \pi^{\pm} K^{+} K^{-}$ \cite{LHCb:2019xmb,LHCb:2019jta,LHCb:2019sus}.
For $B^{\pm} \rightarrow \pi^{\pm} K^{+} K^{-}$, LHCb measured the contribution of the resonance
$K^*(892)$ in the $\pi^{\pm} K^{\mp}$ final state, and the contribution of the resonances
$\rho(1450), \phi(1020), f_2(1270)$ in the $K^{\pm} K^{\mp}$ channel. In this work, we will focus on
the resonant contribution in the $B^-\to K^+K^-\pi^-$ decay.

Using the narrow-width approximation, the resonant contribution of the three-body $B$ decay can be
reduced to various quasi two-body decays, where the $B$ meson firstly decays into a resonance $R$ and
a pion or a kaon $M_1$, $B \to R+M_1$, and subsequently the resonance $R$ decays into the other two
final-state mesons, $R\to M_2+M_3$. 
The latter process simply depends on the strong couplings of
the resonances to the two mesons, while in the first process one has to deal with nontrivial matrix
elements like $\langle R, M_1 | {\cal O} | B\rangle$, with $ {\cal O} $ being some four-quark operator.
In the literature, there are some phenomenological studies of the three-body $B$ decays
\cite{Cheng:2002qu,Cheng:2013dua,Cheng:2020ipp}, where these matrix elements are handled by
naive factorization.  On the other hand,  a more theoretical and widely used method for the
quasi two-body decays of heavy mesons is perturbative QCD (PQCD), for which we refer e.g. to
Refs.~\cite{Li:2016tpn,Li:2017mao,Ma:2017kec,Li:2018lbd,Xing:2019xti,Zou:2020ool,Wang:2018xux,Wang:2020nel}.
In the PQCD approach, the two mesons produced by the resonance are usually described by the
two-meson light-cone distribution amplitudes (LCDAs).  However, in a decay process involving a
number of resonances, one has to search for all the LCDAs corresponding to them. Therefore, an
obvious question is whether these LCDAs are all available or reliable, especially for the
excited states like $\rho(1450)$ or $K_0^*(1430)$. To avoid this problem, one can search for
a method which only depends on the familiar LCDAs like those of $B$ meson or the ground state
pseudoscalar mesons. 

In this work, we will use light-cone sum rules (LCSR) to calculate the matrix elements for the
$B \to R + M_1$ decays. Generally, LCSR are used for the calculation of the form factors of the
semi-leptonic $B$ or $D$ decays. However, recently the use of LCSR for two-body non-leptonic
decays $B/D \to \pi\pi$ has been developed in
Ref.~\cite{Khodjamirian:2000mi,Khodjamirian:2003eq,Khodjamirian:2017zdu}. For the case of $B \to R + M_1$,
one can follow a similar idea to perform the LCSR calculation, where all the non-perturbative
inputs are the LCDAs. The only difference is that here we use the LCDAs of the $B$ meson instead of
the final pseudo-scalar meson. The advantage of LCSR is that in its framework the final states
can be created by two interpolating currents, which have a definite form no matter the state it
creates is the ground state or an excited state. The way to extract the pole contribution of the
excited states is to apply a two-step sum rules, which will be illustrated in the following sections.

This article is organized as follows: In Section~\ref{sec:decayAmps}, we present the form of all the
relevant decay amplitudes in the narrow-width approximation. In Section~\ref{sec:lc_sum_rules},
we introduce the LCSR calculation at the hadron level, which includes the case of scalar,
vector and tensor resonances. In Section~\ref{sec:lc_sum_rules_QCD} we introduce the LCSR calculation
at the quark-gluon level, where we take the case of vector resonance as an example to present the
calculational details. In Section~\ref{sec:phenomenology}, we present the numerical results including
the decay amplitudes and the branching fractions. Section~\ref{sec:conclusion} gives the conclusions
of this work.

\newpage
\section{$B^-\to K^+ K^- \pi^-$ decay amplitude}\label{sec:decayAmps}

In this section, we firstly present the form of the decay amplitude together with the relevant effective
Hamiltonian. From Ref.~\cite{Cheng:2013dua}, the decay amplitude of $B^-\to K^+ K^- \pi^-$ reads 
\begin{equation}
\left\langle K^+ K^- \pi^-\left|\mathcal{H}_{\mathrm{eff}}\right| B^-\right\rangle=\frac{G_{F}}
{\sqrt{2}} \sum_{p=u, c} \lambda_{p}\left\langle K^+ K^- \pi^-\left|H_{p}\right| B^-\right\rangle,
\label{eq:EffecAmp}
\end{equation}
where $\lambda_{p} \equiv V_{p b}^{} V_{p d}^{*}$. The expression of the Hamiltonian $H_p$ is
\begin{align}
H_{p}=& a_{1} \delta_{p u}(\bar{u} b)_{V-A} \otimes(\bar{d} u)_{V-A}+a_{2} \delta_{p u}(\bar{d} b)_{V-A}
\otimes(\bar{u} u)_{V-A}+a_{3}(\bar{d} b)_{V-A} \otimes \sum_{q}(\bar{q} q)_{V-A} \nonumber\\
&+a_{4}^{p} \sum_{q}(\bar{q} b)_{V-A} \otimes(\bar{d} q)_{V-A}+a_{5}(\bar{d} b)_{V-A} \otimes
\sum_{q}(\bar{q} q)_{V+A}  \nonumber\\
&-2 a_{6}^{p} \sum_{q}(\bar{q} b)_{S-P} \otimes(\bar{d} q)_{S+P}+a_{7}(\bar{d} b)_{V-A} \otimes \sum_{q}
\frac{3}{2} e_{q}(\bar{q} q)_{V+A}  \nonumber\\
&-2 a_{8}^{p} \sum_{q}(\bar{q} b)_{S-P} \otimes \frac{3}{2} e_{q}(\bar{d} q)_{S+P}+a_{9}(\bar{d} b)_{V-A}
\otimes \sum_{q} \frac{3}{2} e_{q}(\bar{q} q)_{V-A}  \nonumber\\
&+a_{10}^{p} \sum_{q}(\bar{q} b)_{V-A} \otimes \frac{3}{2} e_{q}(\bar{d} q)_{V-A}.
\end{align}
The notations used here are: $\left(\bar{q} q^{\prime}\right)_{V \pm A} \equiv \bar{q}
\gamma_{\mu}\left(1 \pm \gamma_{5}\right) q^{\prime},\left(\bar{q} q^{\prime}\right)_{S \pm P} \equiv
\bar{q}\left(1 \pm \gamma_{5}\right) q^{\prime}$, and the summation on $q$ runs over $u,d,s$. The
pertinent Wilson coefficients are taken from Ref.~\cite{Cheng:2020ipp}:
\begin{align}
& a_{1} \approx 0.988 \pm 0.102 i, \quad a_{2} \approx 0.183-0.348 i, \quad a_{3} \approx-0.0023+0.0174 i,
\quad a_{5} \approx 0.00644-0.0231 i, \nonumber\\
& a_{4}^{u} \approx-0.025-0.021 i, \quad a_{4}^{c} \approx-0.030-0.012 i, \quad a_{6}^{u} \approx-0.042-0.014 i,
\quad a_{6}^{c} \approx-0.045-0.005 i \nonumber\\
& a_{7} \approx(-0.5+2.7 i) \times 10^{-4}, \quad a_{8}^{u} \approx(5.2-1.0 i) \times 10^{-4}, \quad a_{8}^{c}
\approx(5.0-0.5 i) \times 10^{-4}, \nonumber\\
& a_{9} \approx(-8.9-0.9 i) \times 10^{-3}, \quad a_{10}^{u} \approx(-1.45+3.12 i) \times 10^{-3},
\quad a_{10}^{c} \approx(-1.51+3.17 i) \times 10^{-3} \label{WilsonCoeff}
\end{align}
at the renormalization scale  $\mu=4.18$~GeV. The matrix element of $T_p$ in Eq.~(\ref{eq:EffecAmp})
can be arranged as a sum of the following matrix elements \cite{Cheng:2013dua}:
\begin{align}
\left\langle\pi^{-} K^{+} K^{-}\left|H_{p}\right| B^{-}\right\rangle=&\left\langle K^{+} K^{-}\pi^{-}
\left|{\cal O}_1\right| B^{-}\right\rangle\left[a_{1} \delta_{p u}+a_{4}^{p}+a_{10}^{p}-\left(a_{6}^{p}
+a_{8}^{p}\right) r_{\chi}^{\pi}\right] \nonumber\\
&+\left\langle K^{+} K^{-}\pi^{-}\left|{\cal O}_2\right| B^{-}\right\rangle\left(a_{2} \delta_{p u}+a_{3}
+a_{5}+a_{7}+a_{9}\right) \nonumber\\
&+\left\langle K^{+} K^{-}\pi^{-}\left|{\cal O}_3\right| B^{-}\right\rangle\left[a_{3}+a_{4}^{p}+a_{5}
-\frac{1}{2}\left(a_{7}+a_{9}+a_{10}^{p}\right)\right] \nonumber\\
&+\left\langle K^{+} K^{-}\pi^{-}\left|{\cal O}_4\right| B^{-}\right\rangle\left[a_{3}+a_{5}
-\frac{1}{2}\left(a_{7}+a_{9}\right)\right] \nonumber\\
&+\left\langle K^{+} K^{-}\pi^{-}\left|{\cal O}_5\right| B^{-}\right\rangle\left(-2 a_{6}^{p}+a_{8}^{p}\right)
\nonumber\\
&+\left\langle K^{+} K^{-}\pi^{-}\left|{\cal O}_6\right| B^{-}\right\rangle\left(a_{4}^{p}-\frac{1}{2}
a_{10}^{p}\right) \nonumber\\
&+\left\langle K^{+} K^{-}\pi^{-}\left|{\cal O}_7\right| B^{-}\right\rangle\left(-2 a_{6}^{p}+a_{8}^{p}
\right),
\label{eq:SumOimatrix}
\end{align}
where $r_{\chi}^{\pi}(\mu)=2 {m_{\pi}^{2}}/[{m_{b}(\mu)\left(m_{d}(\mu)-m_{u}(\mu)\right)}]$,
with $\mu=2.1$ GeV, $m_b(\mu)=4.94$~GeV, $m_d(\mu)=5$~MeV and $m_u(\mu)=2.2$~MeV. The four-quark operators
${\cal O}_i$ are
\begin{align}
&{\cal O}_1=(\bar{u}b)_{V-A}(\bar{d}u)_{V-A},~~~~
{\cal O}_2=(\bar{d}b)_{V-A}(\bar{u}u)_{V-A},~~~~
{\cal O}_3=(\bar{d}b)_{V-A}(\bar{d}d)_{V-A},\nonumber\\
&{\cal O}_4=(\bar{d}b)_{V-A}(\bar{s}s)_{V-A},~~~~
{\cal O}_5=(\bar{d}b)(\bar{d}d),~~~~
{\cal O}_6=(\bar{s}b)_{V-A}(\bar{d}s)_{V-A},~~~~{\cal O}_7=(\bar{s}b)(\bar{d}s).
\end{align}
It should be mentioned that in principle Eq.~(\ref{eq:SumOimatrix}) also contain two three-particle terms:
\begin{align}
&\left\langle K^{+} K^{-} \pi^{-}\left|(\bar{d} u)_{V-A}\right| 0\right\rangle\left\langle 0\left|
(\bar{u} b)_{V-A}\right| B^{-}\right\rangle\left(a_{1} \delta_{p u}+a_{4}^{p}+a_{10}^{p}\right) \nonumber\\
+&\left\langle K^{+} K^{-} \pi^{-}\left|\bar{d}\left(1+\gamma_{5}\right) u\right| 0\right\rangle\left
\langle 0\left|\bar{u} \gamma_{5} b\right| B^{-}\right\rangle\left(2 a_{6}^{p}+2 a_{8}^{p}\right).
\end{align}
However, since these three-particle matrix elements are suppressed in the chiral limit \cite{Cheng:2002qu},
we will not consider them in this work.

In general, most of the resonant contribution to the matrix element in Eq.~(\ref{eq:SumOimatrix})
can be described by the quasi-two-body decay process with the use of the  Breit-Wigner (BW) formalism.
However, for the scalar meson $f_0(980)$ with its mass near the $K\bar K$ threshold,  using the BW formalism
is inappropriate. Instead one has to use the \text {Flatté} approach \cite{Flatte:1976xu,Flatte:1976xv}
for such resonances.
Particularly, for the $f_0(980)$ contribution, one has to replace the $m_{f_0(980)}\Gamma_{f_0(980)}$ term
in the standard BW formalism by $g_{\pi\pi}\rho_{\pi\pi}+g_{KK}F_{KK}^2\rho_{KK}$ \cite{Shi:2015kha},
where
\begin{align}
\rho_{\pi \pi}=\frac{2}{3} \sqrt{1-\frac{4 m_{\pi^{\pm}}^{2}}{m_{\pi \pi}^{2}}}+\frac{1}{3} \sqrt{1-\frac{4 m_{\pi^{0}}^{2}}{m_{\pi \pi}^{2}}}, \quad \rho_{K K}=\frac{1}{2} \sqrt{1-\frac{4 m_{K^{\pm}}^{2}}{m_{\pi \pi}^{2}}}+\frac{1}{2} \sqrt{1-\frac{4 m_{K^{0}}^{2}}{m_{\pi \pi}^{2}}}
\end{align}
with $g_{\pi\pi}=167$ MeV and $g_{KK}/g_{\pi\pi}=3.05$ \cite{LHCb:2014ooi}, and $m_{\pi\pi}^2$ is the invariant
mass squared of the two-pion system.
$F_{K K}=\exp \left(-\alpha k^{2}\right)$ is an correction introduced by the LHCb fit above the
$K\bar K$ threshold with $\alpha=2\ {\rm GeV}^{-2}$, and $k$ is the magnitude of the kaon three-momentum in the $K\bar K$ rest frame \cite{LHCb:2014ooi}.
Here, we will simply choose $m_{K^{\pm}}=m_{K^{0}}\equiv m_K$ and $m_{\pi^{\pm}}=m_{\pi^{0}}\equiv m_{\pi}$.
For further discussion, see e.g.  Ref.~\cite{Baru:2004xg}.

The $B$ meson firstly decays into a resonance $R$ and a meson $M_1$, and then the resonance $R$
decays into the rest two mesons $M_2, M_3$. The first process is described by the matrix
elements $\langle R, M_1\left | {\cal O}_i\right|B^-\rangle$, while the second process depends
on the strong coupling of $R$ to $M_2, M_3$. It should be mentioned that in Ref.~\cite{Cheng:2013dua},
naive factorization is used to factorize each matrix element in Eq.~(\ref{eq:SumOimatrix}) into
two current induced terms. For example: 
\begin{equation}
\left\langle K^{+} K^{-}\pi^{-}\left|{\cal O}_1\right| B^{-}\right\rangle=\left\langle K^{+} K^{-}
\left|(\bar{u}b)_{V-A}\right| B^{-}\right\rangle\langle \pi^-\left|(\bar{d}u)_{V-A}\right|0\rangle.
\end{equation}
This means that if the $K^{+} K^{-}$ are produced by a resonance $R$, what one actually calculates
is a factorized form of $\langle R, \pi^-\left | {\cal O}_1\right|B^-\rangle$
\begin{equation}
\langle R, \pi^-\left | {\cal O}_1\right|B^-\rangle=\left\langle R\left|(\bar{u}b)_{V-A}\right|
B^{-}\right\rangle\langle \pi^-\left|(\bar{d}u)_{V-A}\right|0\rangle.\label{naiveFactor}
\end{equation}
The first matrix element on the right-hand side is parameterized by transition form factors,
while the second one is described by the decay constant of the resonance. In this work, with
the use of LCSR, we do not have to assume this factorization,  instead we are able to calculate
the matrix element $\langle R, \pi^-\left | {\cal O}_i\right|B\rangle$ as a whole. 

We follow Ref.~\cite{Cheng:2020ipp} to introduce the resonances appearing in the $B^-\to K^+ K^-\pi^-$ decay.
In terms of the resonant contribution to the ${\cal O}_1$ matrix element, we include a scalar
resonance $f_0(980)$, a vector resonance $\rho(1450)$ and a tensor resonance $f_2(1270)$:
\begin{align}
&\langle\pi^{-}(p_{1})K^{+}(p_{2})K^{-}(p_{3})|{\cal O}_1|B^{-}(p)\rangle_R\nonumber\\
=&\frac{g^{f_{0}(980)\rightarrow K^{+}K^{-}}}{s_{23}-m_{f_{0}(980)}^{2}+i\left(g_{\pi\pi}\rho_{\pi\pi}+g_{KK}F_{KK}^2\rho_{KK}\right)}
\langle f_{0}(980)(p-p_{1})\pi^{-}(p_{1})|{\cal O}_1|B^{-}(p)\rangle\nonumber\\
&+\frac{g^{\rho(1450)\rightarrow K^{+}K^{-}}}{s_{23}-m_{\rho(1450)}^{2}+i\ m_{\rho(1450)}\Gamma_{\rho(1450)}}
\sum_{\lambda}\epsilon(\lambda)\cdot(p_{2}-p_{3})\langle\rho(1450)(p-p_{1},\lambda)\pi^{-}(p_{1})|
{\cal O}_1|B^{-}(p)\rangle\nonumber\\
&+\frac{g^{f_{2}(1270)\rightarrow K^{+}K^{-}}}{s_{23}-m_{f_{2}(1270)}^{2}+i\ m_{f_{2}(1270)}\Gamma_{f_{2}(1270)}}
\sum_{\lambda}\epsilon_{\mu\nu}(\lambda)p_{2}^{\mu}p_{3}^{\nu}\langle f_{2}(1270)(p-p_{1},\lambda)\pi^{-}(p_{1})
|{\cal O}_1|B^{-}(p)\rangle.\label{O1matrix}
\end{align}
The strong coupling constants are defined as in Ref.~\cite{Cheng:2020ipp}. Further, $s_{23}=(p_2+p_3)^2$,
$\epsilon_{\mu}(\lambda)$ denotes the polarization vector of the $\rho(1450)$ and
$\epsilon_{\mu\nu}(\lambda)$ denotes the polarization tensor of the $f_2(1270)$. In terms of
the resonant contribution to the ${\cal O}_{2,3}$ matrix element, we include the vector
resonance $\rho(1450)$:
\begin{align}
&\langle\pi^{-}(p_{1})K^{+}(p_{2})K^{-}(p_{3})|{\cal O}_{2,3}|B^{-}(p)\rangle_R\nonumber\\
=&\frac{g^{\rho(1450)\rightarrow K^{+}K^{-}}}{s_{23}-m_{\rho(1450)}^{2}+i\ m_{\rho(1450)}\Gamma_{\rho(1450)}}
\sum_{\lambda}\epsilon(\lambda)\cdot(p_{2}-p_{3})\langle\rho(1450)(p-p_{1},\lambda)\pi^{-}(p_{1})|
{\cal O}_{2}|B^{-}(p)\rangle.\label{O23matrix}
\end{align}
For the ${\cal O}_{4}$ matrix element, we include the vector resonance $\phi(1020)$:
\begin{align}
&\langle\pi^{-}(p_{1})K^{+}(p_{2})K^{-}(p_{3})|{\cal O}_{4}|B^{-}(p)\rangle_R\nonumber\\
=&\frac{g^{\phi(1020)\rightarrow K^{+}K^{-}}}{s_{23}-m_{\phi(1020)}^{2}+i\ m_{\phi(1020)}\Gamma_{\phi(1020)}}
\sum_{\lambda}\epsilon(\lambda)\cdot(p_{2}-p_{3})\langle\phi(1020)(p-p_{1},\lambda)\pi^{-}(p_{1})|
{\cal O}_{4}|B^{-}(p)\rangle.\label{O4matrix}
\end{align}
For the ${\cal O}_{5}$ matrix element, we include the scalar resonance $f_0(980)$:
\begin{align}
&\langle\pi^{-}(p_{1})K^{+}(p_{2})K^{-}(p_{3})|{\cal O}_{5}|B^{-}(p)\rangle_R\nonumber\\
=&\frac{g^{f_{0}(980)\rightarrow K^{+}K^{-}}}{s_{23}-m_{f_{0}(980)}^{2}+i\left(g_{\pi\pi}\rho_{\pi\pi}+g_{KK}F_{KK}^2\rho_{KK}\right)}\langle
f_{0}(980)(p-p_{1})\pi^{-}(p_{1})|{\cal O}_5|B^{-}(p)\rangle.\label{O5matrix}
\end{align}
For the ${\cal O}_{6}$ matrix element, we include a scalar resonance $K_0^*(1430)$ and a
vector resonance $K^*(892)$:
\begin{align}
&\langle\pi^{-}(p_{1})K^{+}(p_{2})K^{-}(p_{3})|{\cal O}_6|B^{-}(p)\rangle_R\nonumber\\
=&\frac{g^{K^*(892)\rightarrow K^{+}\pi^{-}}}{s_{12}-m_{K^*(892)}^{2}+i\ m_{K^*(892)}\Gamma_{K^*(892)}}\sum_{\lambda}
\epsilon(\lambda)\cdot(p_{1}-p_{2})\langle K^*(892)(p-p_{3},\lambda)K^{-}(p_{3})|{\cal O}_6|B^{-}(p)\rangle
\nonumber\\
&+\frac{g^{K_{0}^*(1430)\rightarrow K^{+}\pi^{-}}}{s_{12}-m_{K_{0}^*(1430)}^{2}+i\ m_{K_{0}^*(1430)}\Gamma_{K_{0}^*(1430)}}
\langle K_{0}^*(1430)(p-p_{3})K^{-}(p_{3})|{\cal O}_6|B^{-}(p)\rangle,\label{O6matrix}
\end{align}
where $s_{12}=(p_1+p_2)^2$. For the ${\cal O}_{7}$ matrix element, we include the scalar resonance
$K_0^*(1430)$:
\begin{align}
&\langle\pi^{-}(p_{1})K^{+}(p_{2})K^{-}(p_{3})|{\cal O}_7|B^{-}(p)\rangle_R\nonumber\\
=&\frac{g^{K_{0}(1430)\rightarrow K^{+}\pi^{-}}}{s_{12}-m_{K_{0}(1430)}^{2}+i\ m_{K_{0}(1430)}\Gamma_{K_{0}(1430)}}
\langle K_{0}(1430)(p-p_{3})K^{-}(p_{3})|{\cal O}_7|B^{-}(p)\rangle.\label{O7matrix}
\end{align}
The matrix elements $\langle R, M\left | {\cal O}_i\right|B^-\rangle$ appearing in
Eqs.~(\ref{O1matrix})-(\ref{O7matrix}) can be parameterized as
\begin{align}
&\langle\rho(1450)(p-p_{1},\lambda)\pi^{-}(p_{1})|{\cal O}_1|B^{-}(p)\rangle=T_{{\cal O}_1}^{\rho(1450)}
\epsilon^*(p-p_1,\lambda)\cdot p_1,\nonumber\\
&\langle f_{0}(980)(p-p_{1})\pi^{-}(p_{1})|{\cal O}_1|B^{-}(p)\rangle=T_{{\cal O}_1}^{f_0(980)},\nonumber\\
&\langle f_2(1270)(p-p_{1},\lambda)\pi^{-}(p_{1})|{\cal O}_1|B^{-}(p)\rangle=T_{{\cal O}_1}^{\rho(1270)}
\epsilon^*_{\mu\nu}(p-p_1,\lambda) p_1^{\mu}p_1^{\nu},\nonumber\\
&\langle\rho(1450)(p-p_{1},\lambda)\pi^{-}(p_{1})|{\cal O}_{2,3}|B^{-}(p)\rangle=
T_{{\cal O}_{2,3}}^{\rho(1450)}\epsilon^*(p-p_1,\lambda)\cdot p_1,\nonumber\\
&\langle\phi(1020)(p-p_{1},\lambda)\pi^{-}(p_{1})|{\cal O}_{4}|B^{-}(p)\rangle=T_{{\cal O}_{4}}^{\phi(1020)}
\epsilon^*(p-p_1,\lambda)\cdot p_1,\nonumber\\
&\langle f_{0}(980)(p-p_{1})\pi^{-}(p_{1})|{\cal O}_5|B^{-}(p)\rangle=T_{{\cal O}_5}^{f_0(980)},\nonumber\\
&\langle K^*(892)(p-p_{3},\lambda)K^{-}(p_{3})|{\cal O}_6|B^{-}(p)\rangle=T_{{\cal O}_6}^{K^*(892)}
\epsilon^*(p-p_3,\lambda)\cdot p_3,\nonumber\\
&\langle K_0^*(1430)(p-p_{3})K^{-}(p_{3})|{\cal O}_{6,7}|B^{-}(p)\rangle=T_{{\cal O}_{6,7}}^{K_0^*(1430)}.
\end{align}
The main task of this work is to obtain these $T_{{\cal O}_{i}}^R$ using the LCSR  approach.

\section{Hadron level calculation in LCSR}
\label{sec:lc_sum_rules}

In this and the next section, we will give an introduction to the calculation of the $B^- \to R, M$
decays within the LCSR approach. In the spirit of LCSR, one should define an appropriate
correlation function corresponding to the process to be studied. Due to the quark-hadron duality,
one has to calculate the correlation function both at the hadron and quark-gluon level.
In this section, we will take the decays induced by ${\cal O}_1$ as an example to introduce
the calculation at the hadron level, while the decays induced by ${\cal O}_{2,\cdots 7}$ can be derived
similarly. 
\subsection{$B^- \to f_0(980) \pi^-$ induced by ${\cal O}_1$}

In the framework of LCSR and with respect to the decay process $B^- \to f_0(980) \pi^-$, we
firstly define a three-point correlation function as
\begin{equation}
\Pi_{\alpha}^{S{\cal O}_1}\left(p,q,k\right)=i^2 \int d^4 x d^4 y e^{i (p-q)\cdot y} e^{i (q+k)\cdot x}
\langle 0|T\left\{j_{5 \alpha}^{\pi}(y){\cal O}_1(0)j_{f_0}(x)\right\}|B(p)\rangle,
\label{eq:CorreFunc}
\end{equation}
where the final pion as well as the S-wave resonance $f_0(980)$ are created by the
interpolating currents  $j_{5 \alpha}^{\pi}$ and $j_{f_0}$, respectively:
\begin{align}
j_{5 \alpha}^{\pi}=\bar u \gamma_{\alpha}\gamma_5 d,~~~~
j_{f_0}=\bar s s\  {\rm cos} \alpha +(\bar u u+\bar d d) \frac{{\rm sin}\alpha}{\sqrt{2}}.
\end{align}
Here, $\alpha$ is the mixing angle between the $f_0(980)$ and $\sigma(500)$, which is chosen as
$\alpha=20^{\circ}$ \cite{Cheng:2020ipp}. In terms of the flavor wave function, such
mixing is given by \cite{Bediaga:2003zh}
\begin{align}
\sigma=  \frac{{\rm cos}\alpha}{\sqrt{2}} (\bar u u+\bar d d) -{\rm sin} \alpha\  \bar s s,~~~~
f_0=\frac{{\rm sin}\alpha}{\sqrt{2}} (\bar u u+\bar d d) +{\rm cos} \alpha\  \bar s s.
\end{align}
In Eq.~(\ref{eq:CorreFunc}), $k$ is an auxiliary momentum which flows into ${\cal O}_1(0)$.
This auxiliary momentum was introduced in Ref.~\cite{Khodjamirian:2000mi}, where the non-leptonic
$B\to \pi\pi$ decay was studied using LCSR. In general, the correlation function in
Eq.~(\ref{eq:CorreFunc}) can be parameterized as
\begin{equation}
\Pi_{\alpha}^{S{\cal O}_1}\left(p,q,k\right)=(p-q)_{\alpha}F_S^{{\cal O}_1}+q_{\alpha}G_{S}^{{\cal O}_1}+k_{\alpha}
H_{S}^{{\cal O}_1}+\epsilon_{\alpha\beta\rho\sigma}p^{\beta}q^{\rho}k^{\sigma}I_{S}^{{\cal O}_1},
\label{eq:CorreFuncPara}
\end{equation}
where the $F_S^{{\cal O}_1}\sim I_S^{{\cal O}_1}$ are functions of the Lorentz invariants
$p^2=m_B^2,~(p-q)^2,~(q+k)^2,~k^2,~q^2$ and ${\bar P}^2=(p-q-k)^2$.
Since the correlation function will be calculated via the Operator-Product-Expansion (OPE), we must
require that the spacetime intervals are small enough: $x^2\sim y^2\sim(x-y)^2\sim 0$. This means
that the external momenta must be extensively spacelike: $(p-q)^2\sim (q+k)^2 \sim {\bar P}^2\ll0$.
On the other hand, as proposed in Ref.~\cite{Khodjamirian:2000mi}, to simplify the calculation, we
can set the remaining squared momenta $k^2$ and $q^2$ to zero.

At the hadronic level, one can firstly insert a complete state with the same quantum number
as $j_{f_0}$ into the correlation function, which then becomes
\begin{equation}
\Pi_{\alpha}^{S{\cal O}_1}\left(p,q,k\right)_{\rm Hadron}=-i \frac{m_{f_0}f_{f_0}}{(q+k)^2-m_{f_0}^2}\int d^4 y
e^{i (p-q)\cdot y}\langle f_0(q+k)|T\{j_{5\alpha}^{\pi}(y){\cal O}_1(0)\}|B(p)\rangle+\cdots,
\label{eq:CorreFuncInsertf0}
\end{equation}
where $f_{f_0}$ is the decay constant of the $f_0(980)$ defined via
$\langle 0|j_{f_0}(0)|f_0(p_f)\rangle = m_{f_0}f_{f_0}$.
We have explicitly kept the contribution from the lowest $f_0(980)$ state, and attributed the
higher multi-particle states and continuous spectrum into the ellipsis. On the other hand, the
same correlation function can be calculated by OPE at the quark-gluon level which is denoted as
$\Pi_{\alpha}^{{\cal O}_1}\left(p,q,k\right)_{\rm QCD}$. It should match with that in
Eq.~(\ref{eq:CorreFuncInsertf0}), which leads to
\begin{align}
&-i \frac{m_{f_0}f_{f_0}}{(q+k)^2-m_{f_0}^2}\int d^4 y e^{i (p-q)\cdot y}\langle f_0(q+k)|T\{j_{5\alpha}^{\pi}(y)
{\cal O}_1(0)\}|B(p)\rangle\nonumber\\
&=\frac{1}{\pi}\int_0^{s_{f_0}}d s_1\frac{{\rm Im}\Pi_{\alpha}^{S{\cal O}_1}\left[(p-q)^2,s_1,\bar P ^2
    \right]_{\rm QCD}}{s_1-(q+k)^2},
\label{eq:CorreFuncInsertf0match1}
\end{align}
where $s_{f_0}$ is the threshold parameter of the sum rule, which is chosen as the mass squared of the
first excited state above the $f_0(980)$ with $J^{PC}=0^{++}$, namely the $f_0(1370)$. On the right-hand side,
the quark-gluon level correlation function is expressed by a dispersion integral. According to the
quark-hadron duality, its integration from $s_{f_0}$ to infinity is canceled by the
contribution of higher excited states on the left-hand side.

The matrix element on the left-hand side of Eq.~(\ref{eq:CorreFuncInsertf0match1}) can be understood
as the emission of a hard momentum $\bar P$ from the $B$ meson which then leaves a $f_0(980)$ as the
final state. According to Eq.~(\ref{eq:CorreFuncInsertf0match1}) and the fact that the right-hand side
is an analytic function of $\bar P^2$, one can perform an analytic continuation to transform the
extensively spacelike $\bar P^2$ to the timelike region. Note that $k$ as an auxiliary momentum will
be set to zero so that $\bar P \to p-q$ at last. Since $p-q$ corresponds to the momentum flow
out off the pion vertex, it is natural to choose $\bar P^2=m_{\pi}^2$. At this physical region,
the outgoing $f_0$ can be moved to the initial state with an inverse momentum
\begin{align}
&-i \frac{m_{f_0}f_{f_0}}{(q+k)^2-m_{f_0}^2}\int d^4 y e^{i (p-q)\cdot y}\langle 0|T\{j_{5\alpha}^{\pi}(y)
{\cal O}_1(0)\}|B(p)f_0(-q-k)\rangle\nonumber\\
&=\frac{1}{\pi}\int_0^{s_{f_0}}d s_1\frac{{\rm Im}\Pi_{\alpha}^{S{\cal O}_1}\left[(p-q)^2,s_1,\bar P ^2=m_{\pi}^2\right]_{\rm QCD}}{s_1-(q+k)^2}.
\label{eq:CorreFuncInsertf0match2}
\end{align}
Then, by inserting a complete state with the same quantum number as $j_{5\alpha}^{\pi}$ on the left-hand side,
using the quark-hadron duality to cancel out its higher excited contributions, and using crossing symmetry
to move the $f_0$ to the final-state again, one arrives at
\begin{align}
&i \frac{m_{f_0}f_{f_0}f_{\pi}}{((q+k)^2-m_{f_0}^2)((p-q)^2-m_{\pi}^2)}(p-q)_{\alpha}\langle f_0(q+k)\pi(p-q)|
{\cal O}_1(0)|B(p)\rangle\nonumber\\
&=\frac{1}{\pi^2}\int_0^{s_{f_0}}d s_1\int_0^{s_{\pi}}d s_2\frac{{\rm Im}^2\Pi_{\alpha}^{S{\cal O}_1}
  \left[s_2,s_1,\bar P ^2=m_{\pi}^2\right]_{\rm QCD}}{(s_1-(q+k)^2)(s_2-(p-q)^2)},
\label{eq:CorreFuncInsertf0match2}
\end{align}
where we have used the definition of the pion decay constant, $\langle 0\left |j_{5\alpha}^{\pi}(0)\right
|\pi(p)\rangle=i f_{\pi}p_{\alpha}$, and $s_{\pi}$ is chosen as the squared mass of the $\pi(1300)$.
The symbol ${\rm Im}^2$ means extracting the double imaginary part corresponding to the discontinuities
across $s_1$ and $s_2$. Using the Borel transformation formula: 
\begin{equation}
{\cal B}_{p^2}\left[\frac{1}{m^2-p^2}\right]=e^{-m^2/M^2}
\end{equation}
for the external momenta squared $(q+k)^2$ and $(p-q)^2$ in Eq.~(\ref{eq:CorreFuncInsertf0match2}),
and denoting the corresponding Borel parameters as $M_1$ and $M_2$, one arrives at
\begin{align}
&\langle f_0(q+k)\pi(p-q)|{\cal O}_1(0)|B(p)=-\frac{i}{\pi^2 m_{f_0}f_{f_0}f_{\pi}}e^{m_{\pi}^2/M_2^2}
e^{m_{f_0}^2/M_1^2}\nonumber\\
&\times\int_0^{s_{f_0}}d s_1\int_0^{s_{\pi}}d s_2 e^{-s_2/M_2^2}e^{-s1/M_1^2}{\rm Im}^2 F_S^{{\cal O}_1}
\left[s_2,s_1,\bar P ^2=m_{\pi}^2\right]_{\rm QCD}~.
\label{eq:CorreFuncInsertf0matchBorel}
\end{align}
Note that on the left-hand side of Eq.~(\ref{eq:CorreFuncInsertf0match2}), only the Lorentz structure
of $(p-q)_{\alpha}$ appears. Thus only the $F_S^{{\cal O}_1}$ in Eq.~(\ref{eq:CorreFuncPara})  contributes. 

\subsection{$B^- \to \rho(1450) \pi^-$ induced by ${\cal O}_1$}

In this subsection we consider the process $B^- \to \rho(1450) \pi^-$, where the resonance is
a vector particle. The definition of the correlation function is similar except that
the interpolating current for the resonance is changed to a vector current, which is given by
\begin{equation}
\Pi_{\alpha\beta}^{V{\cal O}_1}\left(p,q,k\right)=i^2 \int d^4 x d^4 y e^{i (p-q)\cdot y} e^{i (q+k)\cdot x}\langle
0|T\left\{j_{5 \alpha}^{\pi}(y){\cal O}_1(0)j_{\beta}^V(x)\right\}|B(p)\rangle,
\label{eq:CorreFuncVec}
\end{equation}
with
\begin{equation}
j_{\beta}^V=\frac{1}{\sqrt 2}\left(\bar u \gamma_\beta u-\bar d \gamma_\beta d\right).
\end{equation}
The procedure of inserting the states with the same quantum numbers of $\pi^-$ and $\rho(1450)$
is similar to that of $B^- \to f_0(980) \pi^-$.  However, it should be noted that the $\rho(1450)$
is the second excited state with $J^{PC}=1^{--}$, while the lowest one is in fact the $\rho(770)$.
The strategy to extract the contribution from the $\rho(1450)$ is to construct a two-step sum rule.
First, we just keep the lowest state $\rho(770)$ at the hadron level and attribute the
$\rho(1450)$ to the higher excited spectrum. Then we can calculate the matrix element
$\langle \rho(770)\pi^-\left |{\cal O}_1\right |B^-\rangle$ with the threshold parameter
chosen as $s_{\rho(770)}=m_{\rho(1450)}^2$. At the second step, both $\rho(770)$ and $\rho(1450)$ are
kept and the higher excited spectrum begins from the squared mass of the $\rho(1570)$,
in other words $s_{\rho(1450)}=m_{\rho(1570)}^2$.

In the first step, we obtain the sum rule equation for the $\rho(770)$ as
\begin{align}
&i \frac{m_{\rho(770)}f_{\rho(770)}f_{\pi}(p-q)_{\alpha}}{\left((q+k)^2-m_{\rho(770)}^2\right)\left((p-q)^2-m_{\pi}^2
\right)}\sum_{\lambda}{\epsilon}_{\beta}(q+k,\lambda)\langle \rho(770)(q+k,\lambda)\pi(p-q)|{\cal O}_1(0)|
B(p)\rangle\nonumber\\
&=\frac{1}{\pi^2}\int_0^{s_{\rho(770)}}d s_1\int_0^{s_{\pi}}d s_2\frac{{\rm Im}^2\Pi_{\alpha\beta}^{V{\cal O}_1}
  \left[s_2,s_1,\bar P ^2=m_{\pi}^2\right]_{\rm QCD}}{(s_1-(q+k)^2)(s_2-(p-q)^2)},
\label{eq:CorreFuncInsertRho770a}
\end{align}
where $f_{\rho(770)}$ is defined via
\begin{equation}
\langle 0\left|j_{\beta}^V(0)\right |\rho(770)(p,\lambda)\rangle=m_{\rho(770)}f_{\rho(770)}\epsilon_{\beta}
(p,\lambda), \label{decayConstant770}
\end{equation}
and similarly for the $f_{\rho(1450)}$. On the left-hand side of Eq.~(\ref{eq:CorreFuncInsertRho770a}),
the matrix element not only depends on $p, q$, but also on the auxiliary momentum $k$. Thus
defining $p_{\pi}=p-q,~p_V=q+k$ and using the constraint $\epsilon\cdot p_{\pi}=0$,
it should be parameterized by two terms:
\begin{equation}
\langle \rho(770)(p_V,\lambda)\pi(p_{\pi})|{\cal O}_1(0)|B(p)\rangle=T_{{\cal O}_1}^{\rho(770)}
\epsilon^*(p_V,\lambda)\cdot p_{\pi}+T_{{\cal O}_1}^{\rho(770)\prime} \epsilon^*(p_V,\lambda)\cdot k\ .
\label{Btorho770piPars}
\end{equation}
Although the second term vanishes when we set $k\to 0$ at the end of the calculation, both of them
should be kept in the intermediate steps. Now we need two equations to solve for $T_{{\cal O}_1}^{\rho(770)}$
and $T_{{\cal O}_1}^{\rho(770)\prime}$. To do this, we can contract with two independent vectors,
$\Gamma_i^{\beta}=\{m_B v^{\beta}, q^{\beta}\}$ ($i=1,2$), 
respectively on the both sides of Eq.~(\ref{eq:CorreFuncInsertRho770a}). This leaves only one
Lorentz index for the correlation function on the right-hand side, which again has the same
structures as Eq.~(\ref{eq:CorreFuncPara}):
\begin{equation}
\Gamma_i^{\beta}\Pi_{\alpha\beta}^{V{\cal O}_1}\left(p,q,k\right)=(p-q)_{\alpha}F_{V(i)}^{{\cal O}_1}+q_{\alpha}
G_{V(i)}^{{\cal O}_1}+k_{\alpha}H_{V(i)}^{{\cal O}_1}+\epsilon_{\alpha\beta\rho\sigma}p^{\beta}q^{\rho}k^{\sigma}
I_{V(i)}^{{\cal O}_1}.
\label{eq:CorreFuncParaVec}
\end{equation}
As a result, the sum rule equation becomes
\begin{align}
&i \frac{m_{\rho(770)}f_{\rho(770)}f_{\pi}}{\left(p_V^2-m_{\rho(770)}^2\right)\left(p_{\pi}^2-m_{\pi}^2\right)}
\Gamma_{i\mu}\left[-g^{\mu\nu}+\frac{p_{V}^{\mu}p_{V}^{\nu}}{m_{\rho(770)}^2}\right]\left(T_{{\cal O}_1}^{\rho(770)}
p_{\pi\nu}+T_{{\cal O}_1}^{\rho(770)\prime} k_{\nu}\right)\nonumber\\
&=\frac{1}{\pi^2}\int_0^{s_{\rho(770)}}d s_1\int_0^{s_{\pi}}d s_2\frac{{\rm Im}^2F_{V(i)}^{{\cal O}_1}
\left[s_2,s_1,\bar P ^2=m_{\pi}^2\right]_{\rm QCD}}{(s_1-p_V^2)(s_2-p_{\pi}^2)},
\label{eq:CorreFuncInsertRho770b}
\end{align}
where have used the polarization summation formula for $\epsilon_{\mu}(p_V,\lambda)$
\begin{equation}
\sum_{\lambda}\epsilon^{\mu}(p_V,\lambda)\epsilon^{*\nu}(p_V,\lambda)=-g^{\mu\nu}
+\frac{p_{V}^{\mu}p_{V}^{\nu}}{m_{\rho(770)}^2}.
\end{equation}
All the momentum contractions on the left-hand side of Eq.~(\ref{eq:CorreFuncInsertRho770b}) should be
expressed by the squared momenta, namley $p_V^2, p_{\pi}^2, q^2=0, k^2=0$ and $\bar P^2=m_{\pi}^2$. Note
that the Borel transformation for any polynomial of $p_V^2$ and $p_{\pi}^2$ is zero. Thus we have
to use the trick  $p_V^2=-(m_{\rho(770)}^2-p_V^2)+m_{\rho(770)}^2$ and $p_{\pi}^2=-(m_{\pi}^2-p_V^2)+m_{\pi}^2$
to express the the left-hand side of Eq.~(\ref{eq:CorreFuncInsertRho770b}) as 
\begin{equation}
\frac{C_1}{\left(m_{\pi}^2-p_{\pi}^2\right)(m_{\rho(770)}^2-p_V^2)}+\frac{C_2}{(m_{\pi}^2-p_{\pi}^2)}
+\frac{C_3}{(m_{\rho(770)}^2-p_V^2)}+C_4,
\end{equation}
where the $C_i$ are independent of $p_{\pi}^2$ and $p_V^2$. After the double Borel transformation
in terms of $p_V^2$ and $p_{\pi}^2$, the the $C_{2,3,4}$ terms vanish. Then we  obtain
\begin{align}
T_{{\cal O}_1}^{\rho(770)}=&-\int_0^{s_{\rho(770)}}d s_1\int_0^{s_{\pi}}d s_2\frac{2i\ m_V e^{(m_V^2-s_1)/M_1^2}
e^{(m_{\pi}^2-s_2)/M_2^2}}{\pi^2 f_{\pi}f_V(m_B^4-2 m_B^2 m_{\pi}^2-2 m_B^2 m_V^2+m_{\pi}^4+m_V^4)}\nonumber\\
&\times\left\{{\rm Im}^2F_{V(1)}^{{\cal O}_1}\left[s_2,s_1,m_{\pi}^2\right]_{\rm QCD}+\left(\frac{m_B^2}{m_V^2}
-\frac{m_{\pi}^2}{m_V^2}-1\right){\rm Im}^2F_{V(2)}^{{\cal O}_1}\left[s_2,s_1,m_{\pi}^2\right]_{\rm QCD}\right\},
\label{TO1rho770}\\
T_{{\cal O}_1}^{\rho(770)\prime}=&-\int_0^{s_{\rho(770)}}d s_1\int_0^{s_{\pi}}d s_2\frac{2i\ m_V e^{(m_{\rho(770)}^2-s_1)/M_1^2}
e^{(m_{\pi}^2-s_2)/M_2^2}}{\pi^2 f_{\pi}f_V(m_B^4-2 m_B^2 m_{\pi}^2-2 m_B^2 m_V^2+m_{\pi}^4+m_V^4)}\nonumber\\
&\times\left\{\left(-\frac{m_B^2}{m_V^2}+\frac{m_{\pi}^2}{m_V^2}+2\right){\rm Im}^2F_{V(1)}^{{\cal O}_1}
\left[s_2,s_1,m_{\pi}^2\right]_{\rm QCD}\right.\nonumber\\
&\left.+\left(\frac{m_B^4}{m_V^4}-2 \frac{m_B^2 m_{\pi}^2}{m_V^4}-\frac{m_B^2}{m_V^2}+\frac{m_{\pi}^4}{m_V^4}
-3 \frac{m_{\pi}^2}{m_V^2}\right){\rm Im}^2F_{V(2)}^{{\cal O}_1}\left[s_2,s_1,m_{\pi}^2\right]_{\rm QCD}\right\}.
\label{TO1rho770prime}
\end{align}
We have denoted the mass and decay constant of the $\rho(770)$ by $m_V$ and $f_V$, in order. 

In the second step, we keep both the $\rho(770)$ and the $\rho(1450)$, then similarly to
Eq.~(\ref{eq:CorreFuncInsertRho770a}), the sum rule equation becomes
\begin{align}
&i \frac{m_{\rho(770)}f_{\rho(770)}f_{\pi}(p-q)_{\alpha}}{\left((q+k)^2-m_{\rho(770)}^2\right)
\left((p-q)^2-m_{\pi}^2\right)}\sum_{\lambda}{\epsilon}_{\beta}(q+k,\lambda)\langle \rho(770)(q+k,\lambda)
\pi(p-q)|{\cal O}_1(0)|B(p)\rangle\nonumber\\
&+i \frac{m_{\rho(1450)}f_{\rho(1450)}f_{\pi}(p-q)_{\alpha}}{\left((q+k)^2-m_{\rho(1450)}^2\right)\left((p-q)^2
-m_{\pi}^2\right)}\sum_{\lambda}{\epsilon}_{\beta}^{\prime}(q+k,\lambda)\langle \rho(1450)(q+k,\lambda)\pi(p-q)
|{\cal O}_1(0)|B(p)\rangle\nonumber\\
&=\frac{1}{\pi^2}\int_0^{s_{\rho(1450)}}d s_1\int_0^{s_{\pi}}d s_2\frac{{\rm Im}^2\Pi_{\alpha\beta}^{V{\cal O}_1}
\left[s_2,s_1,\bar P ^2=m_{\pi}^2\right]_{\rm QCD}}{(s_1-(q+k)^2)(s_2-(p-q)^2)}.
\label{eq:CorreFuncInsertRho1450a}
\end{align}
Note that now the upper limit of the $s_1$ integration is $s_{\rho(1450)}=m_{\rho(1570)}^2$. The
parameterization for the second matrix element above is similar to Eq.~(\ref{Btorho770piPars}).
Using the results from Eq.~(\ref{TO1rho770}) and Eq.~(\ref{TO1rho770prime}), we can obtain the
expression for $T_{{\cal O}_1}^{\rho(1450)}$:
\begin{eqnarray}
T_{{\cal O}_1}^{\rho(1450)}=&-&\frac{f_{V}e^{\left(m_{V}^{\prime 2}-m_{V}^{2}\right)/ M_{1}^{2}}\ T_{{\cal O}_1}^{\rho(770)}}
{2f_{V}^{\prime}m_{V}m_{V}^{\prime}\left(m_{B}^{4}-2m_{B}
^{2}\left(m_{\pi}^{2}+m_{V}^{\prime}{}^{2}\right)+m_{\pi}^{4}+m_{V}^{\prime}{}^{4}\right)}\nonumber\\
&&\times\left[m_{B}^{4}\left(m_{V}^{2}+m_{V}^{\prime}{}^{2}\right)-2m_{B}^{2}\left(m_{\pi}^{2}+m_{V}^{2}\right)
\left(m_{V}^{2}+m_{V}^{\prime}{}^{2}\right)+m_{\pi}^{4}\left(m_{V}^{2}+m_{V}^{\prime}{}^{2}\right)\right.\nonumber\\
&&\left.+2m_{\pi}^{2}\left(m_{V}^{4}-m_{V}^{2}m_{V}^{\prime}{}^{2}\right)+2m_{V}^{4}m_{V}^{\prime}{}^{2}\right]
\nonumber\\
&+&\frac{f_{V}m_{V}e^{\left(m_{V}^{\prime 2}-m_{V}^{2}\right)/ M_{1}^{2}}\left(m_{B}^{2}-m_{\pi}^{2}\right)\left(m_{V}^{2}
-m_{V}^{\prime}{}^{2}\right)}{2f_{V}^{\prime}m_{V}^{\prime}\left(m_{B}^{4}-2m_{B}^{2}\left(m_{\pi}^{2}+m_{V}^{\prime}{}^{2}
\right)+m_{\pi}^{4}+m_{V}^{\prime}{}^{4}\right)}T_{{\cal O}_1}^{\rho(770)\prime}\nonumber\\
&-&\int_0^{s_{\rho(1450)}}d s_1\int_0^{s_{\pi}}d s_2\frac{2i\ m_{V}^{\prime}e^{(m_V^{\prime 2}-s_1)/M_1^2}e^{(m_{\pi}^2-s_2)/M_2^2}}
{\pi^{2}f_{\pi}f_{V}^{\prime}\left(m_{B}^{4}-2m_{B}^{2}m_{\pi}^{2}-2m_{B}^{2}m_{V}^{\prime}{}^{2}+m_{\pi}^{4}
+m_{V}^{\prime}{}^{4}\right)}\nonumber\\
&&\times\left\{{\rm Im}^2F_{V(1)}^{{\cal O}_1}\left[s_2,s_1,m_{\pi}^2\right]_{\rm QCD}+\left(\frac{m_B^2}
{m_V^{\prime 2}}-\frac{m_{\pi}^2}{m_V^{\prime 2}}-1\right){\rm Im}^2F_{V(2)}^{{\cal O}_1}\left[s_2,s_1,m_{\pi}^2
\right]_{\rm QCD}\right\},\nonumber\\
\end{eqnarray}
where $m_V^{\prime}$ and $f_V^{\prime}$ denote the mass and decay constant of the $\rho(1450)$, respectiverly.
We do not give the expression for $T_{{\cal O}_1}^{\rho(1450)\prime}$ since it is irrelevant when $k\to 0$ at the
end.

\subsection{$B^- \to f_2(1270) \pi^-$ induced by ${\cal O}_1$}

In this subsection we consider the process of $B^- \to f_2(1270) \pi^-$, where the resonance is
a tensor particle. Note that after  introducing the auxiliary momentum $k$, the $B^- \to f_2(1270) \pi^-$
matrix element should be parameterized by three terms:
\begin{align}
&\langle f_2(1270)(p_T,\lambda)\pi(p_{\pi})|{\cal O}_1(0)|B(p)\rangle\nonumber\\
=&\  \epsilon_{\mu\nu}^*(p_T,\lambda)\left\{T_{{\cal O}_1}^{f_2(1270)} p_{\pi}^{\mu}p_{\pi}^{\nu}+\frac{1}{2}
T_{{\cal O}_1}^{f_2(1270)\prime}  \left(p_{\pi}^{\mu}k^{\nu}+p_{\pi}^{\nu}k^{\mu}\right)+T_{{\cal O}_1}^{f_2(1270)\prime\prime}
k^{\mu}k^{\nu}\right\},\label{Btof21270piPars}
\end{align}
where $p_T=q+k$ and $p_{\pi}=p-q$, and the constraints $\epsilon_{\mu\nu}=\epsilon_{\nu\mu}$,
$\epsilon_{\mu\nu}p_T^{\mu}=0$ are used. Accordingly, when defining the correlation function, we
have to introduce three projection tensors: 
\begin{equation}
\Gamma_{i}^{\mu\nu}=\left\{m_B^2 v^{\mu}v^{\nu},\ \frac{1}{2}m_B (v^{\mu}q^{\nu}+v^{\nu}q^{\mu}),\ q^{\mu}q^{\nu}\right\}
~(i=1,2,3)
\end{equation}
to contract with the tensor current. The projected correlation function then reads  
\begin{equation}
\Gamma_i^{\beta\sigma}\Pi_{\alpha\beta\sigma}^{T{\cal O}_1}\left(p,q,k\right)=i^2 \int d^4 x d^4 y e^{i (p-q)\cdot y}
e^{i (q+k)\cdot x}\Gamma_i^{\beta\sigma}\langle 0|T\left\{j_{5 \alpha}^{\pi}(y){\cal O}_1(0)j_{\beta\sigma}^T(x)
\right\}|B(p)\rangle,
\label{eq:CorreFuncTens}
\end{equation}
where 
\begin{equation}
j_{\beta\sigma}^T=\frac{i}{4\sqrt{2}}\left(\bar u\gamma_{\beta}\stackrel{\leftrightarrow}{\partial}_{\sigma} u
+\bar u\gamma_{\sigma}\stackrel{\leftrightarrow}{\partial}_{\beta} u\right)-(u\leftrightarrow d)
\label{eq:TensorCurrent}
\end{equation}
with $\stackrel{\leftrightarrow}{\partial}=\stackrel{\rightarrow}{\partial}-\stackrel{\leftarrow}{\partial}$
\cite{Aliev:1981ju}. Its parameterization is the same as that of Eq.~(\ref{eq:CorreFuncPara}) and
Eq.~(\ref{eq:CorreFuncParaVec}) :
\begin{equation}
\Gamma_i^{\beta\sigma}\Pi_{\alpha\beta\sigma}^{T{\cal O}_1}\left(p,q,k\right)=(p-q)_{\alpha}F_{T(i)}^{{\cal O}_1}
+q_{\alpha}G_{T(i)}^{{\cal O}_1}+k_{\alpha}H_{T(i)}^{{\cal O}_1}+\epsilon_{\alpha\beta\rho\sigma}p^{\beta}q^{\rho}k^{\sigma}
I_{T(i)}^{{\cal O}_1}.\label{eq:CorreFuncParaTens}
\end{equation}

Then we follow the similar procedure to insert a state with the same quantum numbers of the $f_2(1270)$.
Using the definition of the $f_2(1270)$ decay constant:
\begin{equation}
\langle 0\left|j_{\mu\nu}^V(0)\right | f_2(1270)(p,\lambda)\rangle=m_{f_2(1270)}^2 f_{f_2(1270)}
\epsilon_{\mu\nu}(p,\lambda),\label{decayConstant1270}
\end{equation}
as well as the summation formula of the polarization tensor:
\begin{align}
\sum_{\lambda}\epsilon_{\mu\nu}(p_T,\lambda)\epsilon_{\rho\sigma}^*(p_T,\lambda)=\frac{1}{2}P^T_{\mu\rho}
P^T_{\nu\sigma}+\frac{1}{2}P^T_{\mu\sigma}P^T_{\nu\rho}-\frac{1}{3}P^T_{\mu\nu}P^T_{\rho\sigma}
\end{align}
with $P^T_{\mu\nu}=g_{\mu\nu}-p_{T\mu}p_{T\nu}/m_{f_2(1270)}^2$, we obtain
\begin{align}
T_{{\cal O}_1}^{f_2(1270)}=&\frac{4i}{\pi^2 f_{\pi}f_T m_B^6}e^{m_V^2/M_1^2}e^{m_{\pi}^2/M_2^2}\int_0^{s_{f_2(1270)}}
d s_1\int_0^{s_{\pi}}d s_2 \ \nonumber\\
&\times\sum_{i=1}^3\frac{1}{\xi_i(\lambda_{\pi},\lambda_{T})}e^{-s_1/M_1^2}e^{-s_2/M_2^2}{\rm Im}^2
F_{T(i)}^{{\cal O}_1}\left[s_2,s_1,m_{\pi}^2\right]_{\rm QCD}
\end{align}
where $\lambda_{\pi}=m_{\pi}^2/m_B^2,\ \lambda_{T}=m_{T}^2/m_B^2$. The expression for $\xi_i(\lambda_{\pi},
\lambda_{T})$ can be found in Appendix~A. The threshold is chosen as $s_{f_2(1270)}=m_{f_2^{\prime}(1520)}^2$.
The $T_{{\cal O}_1}^{f_2(1270)\prime}$ and $T_{{\cal O}_1}^{f_2(1270)\prime\prime}$ are irrelevant when
taking $k\to 0$ so that they are not shown here.

\section{Quark-gluon level calculation in LCSR}
\label{sec:lc_sum_rules_QCD}

In this section we will present the calculation of the correlation functions given in
Eq.~(\ref{eq:CorreFunc}), Eq.~(\ref{eq:CorreFuncVec}) and Eq.~(\ref{eq:CorreFuncTens}) on the
quark-gluon level. Since the calculation procedure of these three cases is similar, we only take
Eq.~(\ref{eq:CorreFuncVec}) as an example to give a detailed derivation. 

\begin{figure}
\begin{center}
\includegraphics[width=0.95\columnwidth]{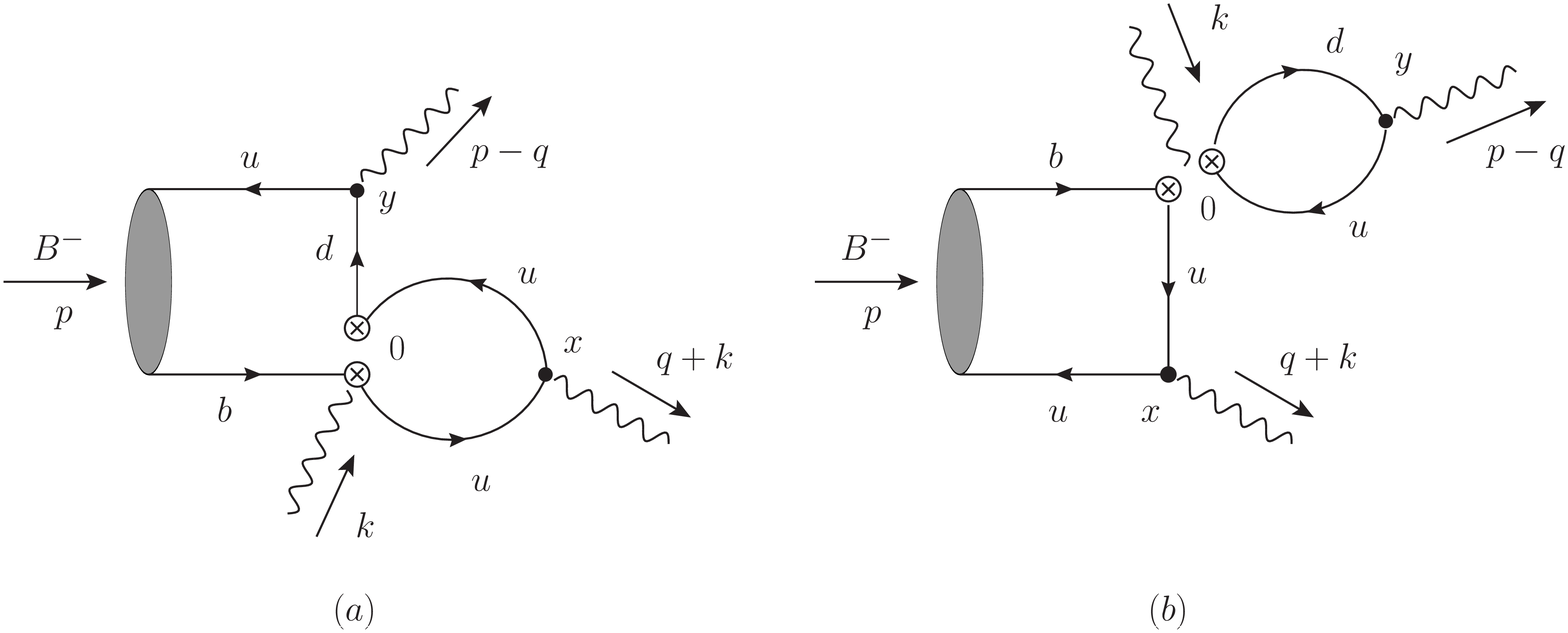} 
\caption{Feynman diagrams of the correlation function given in Eq.~(\ref{eq:CorreFuncVec}).  (a) is
the $W$-exchange diagram, and (b) is the $W$-emission diagram. The gray bubble denotes the $B$ meson,
the pairs of crossed circles represent the four-quark operators ${\cal O}_1$, the black dots 
denote the interpolating currents for final particles and the wiggly lines denote the incoming or outgoing momentum flows.} 
\label{fig:FeynmanDiag} 
\end{center}
\end{figure}
There are two diagrams contributing to the correlation function. As shown in Fig.~\ref{fig:FeynmanDiag},
(a) is the $W$-exchange diagram, and (b) is the $W$-emission diagram. Note that the factorization
given in Eq.~(\ref{naiveFactor}) only involves the contribution from $W$-emission, while it
misses the $W$-exchange effect. We will give a detailed calculation for the diagram (a) within
the light-cone expansion, the calculation for the diagram (b) is similar and will more be given
in any detail here. The definition of the
leading twist light-cone distribution amplitude (LCDA) is
\cite{Grozin:1996pq,Beneke:2000wa,Khodjamirian:2006st,Braun:2017liq}:
\begin{align}
&\left\langle 0\left|\bar{u}_{\alpha}(x)[x, 0] b_{v \beta}(0)\right| \bar{B}_{v}^{0}\right\rangle \nonumber\\
=&-\frac{i f_{B} m_{B}}{4} \int_{0}^{\infty} d \omega e^{-i \omega v \cdot x}\left\{(1+\slashed v)
\left(\phi_{+}^{B}(\omega)+\frac{\phi_{-}^{B}(\omega)-\phi_{+}^{B}(\omega)}{2 v \cdot x} \slashed x\right)
\gamma_{5}\right\}_{\beta \alpha}.\label{BLCDAs}
\end{align}
According to the experience from Ref.~\cite{Shi:2021bvy}, the contribution from the next-to-leading
order twist LCDAs is expected to be one order smaller than that of the leading twist LCDAs, so we do
not consider them in this work.  Note that the first term above has no terms $\sim\slashed x$ and
$\sim v\cdot x$,
thus its contribution to the correlation function can be derived directly as
\begin{align}
&\Gamma_{i}^{\beta}\Pi_{\alpha\beta}^{{\cal O}_{1}}(p,q,k)_{(a)x^0} & \nonumber\\
=&\ \frac{f_{B}m_{B}}{4\sqrt{2}}\int_{0}^{\infty}d\omega\ \phi_{+}^{B}(\omega)\frac{1}{(p_{\pi}-\omega v)^{2}
-m^{2}}\int\frac{d^{4}k_l}{(2\pi)^{4}}\frac{1}{k_{l}^{2}-m^{2}}\frac{1}{(k_l+p_{V})^{2}-m^{2}} \nonumber\\
& \times{\rm tr}\left[\gamma_{\alpha}\gamma_{5}(\slashed p_{\pi}-\omega\slashed v+m)\gamma^{\mu}(1-\gamma_{5})
(\slashed k_l+m)\Gamma_{i}\cdot \gamma(\slashed p_{V}+\slashed k_l+m)\gamma_{\mu}(1-\gamma_{5})(1+\slashed v)
  \gamma_5\right],
\label{Correx0}
\end{align}
where $m$ denotes the light ($u,d$) quark mass which will be approximated as zero in this work.
The double imaginary
part ${\rm Im}^2\equiv{\rm Im}_{p_{\pi}^2}{\rm Im}_{p_{V}^2}$ of the correlation function is proportional
to the discontinuity in terms of the invariants $p_V^2$ and $p_{\pi}^2$, and the discontinuity only
comes from the denominators of the integrand in Eq.~(\ref{Correx0}). It is found that $p_V^2$ and
$p_{\pi}^2$ do not appear in the same denominator, so we can extract the two imaginary parts independently.
The imaginary part in terms of $p_V^2$ can be obtained by using the cutting-rule for the bubble in the
lower right corner of diagram (a). Thus the loop integration of $d^4 k_l$ in this bubble is replaced
by a two-body phase space integration after using the cutting-rule. The tensor basis used for the
two-body phase space integration can be found in  Appendix~B. On the other hand, the imaginary part in
terms of $p_{\pi}^2$ only comes from the first denominator:
\begin{equation}
{\rm Im}_{p_{\pi}^2}\frac{1}{(p_{\pi}-\omega v)^{2}-m^{2}}=(-\pi)\delta\left[\left(1-\frac{\omega}{m_B}\right)
p_{\pi}^2-\omega m_B+\omega^2-m^2\right].
\end{equation}
After calculating the double imaginary part of Eq.~(\ref{Correx0}), and extracting the coefficient of
$p_{\pi\alpha}$, we obtain the corresponding contribution to the imaginary part of $F_{V(i)}^{{\cal O}_1}$:
\begin{align}
{\rm Im}^2F_{V(i)}^{{\cal O}_1}(p_{\pi}^2, p_{V}^2)_{(a)x^0}=&-i \sqrt{2}\pi^5 f_B m_B\int_0^{\infty}d\omega\
\phi_{+}^{B}(\omega)\delta\left[\left(1-\frac{\omega}{m_B}\right)p_{\pi}^2-\omega m_B+\omega^2-m^2\right]
\nonumber\\&\times\int d\Phi_2(p_V^2,k_2,k_3)\ {\cal M}^{V(i)}_{(a)}(p_{\pi},k_2,k_3),
\label{Correx0ImpV}
\end{align}
where we have used the requirements $k^2=q^2=0$ and $\bar P^2=m_{\pi}^2$ as argued in the last section.
Further, $k_2, k_3$ are the momenta of the two light quarks in the bubble. The expression of
${\cal M}^{(i)}_{(a)x^0}(p_{\pi},k_2,k_3)$  can be found in Appendix~C. The two-body phase space integration
is defined as
\begin{equation}
\int d\Phi_2(p_V^2,k_2,k_3)=\int \frac{d^3 \vec{k}_2}{(2\pi)^3}\frac{1}{2E_{k_2}}\frac{d^3 \vec{k}_3}{(2\pi)^3}\frac{1}{2E_{k_3}}\delta^{(4)}(p_V-k_2-k_3)\delta(k_2^2-m^2)\delta(k_3^2-m^2).
\end{equation}
The result of the two-body phase space integration in Eq.~(\ref{Correx0ImpV}) is given in Appendix~B.  

For the term proportional to $\slashed x/v\cdot x$ in Eq.~(\ref{BLCDAs}), we use the following trick
to remove the troublesome denominator $v\cdot x$. We define:
\begin{equation}
\tilde{\phi}_{\pm}^{B}(\omega)=\int_{0}^{\omega}d\tau\ \phi_{\pm}^{B}(\tau)~~\longrightarrow~\phi_{\pm}^{B}(\omega)
=\frac{d}{d\omega}\tilde{\phi}_{\pm}^{B}(\omega)
\end{equation}
so that the integration on the spacetime coordinate can be simplified as
\begin{equation}
\int_{0}^{\infty}d\omega\ e^{-i\omega v\cdot x}\frac{\phi_{\pm}^{B}(\omega)}{v\cdot x}\cdots=i \int_{0}^{\infty}d\omega\
e^{-i\omega v\cdot x}\tilde{\phi}_{\pm}^{B}(\omega)\cdots.
\end{equation}
Then, the only $x$ dependence comes from the term $\sim{\rm exp}[i(q+k)\cdot x]$ in the definition of
the correlation function in Eq.~(\ref{eq:CorreFuncVec}). Thus in the momentum space the term $\sim\slashed x$
can be replaced by $-i\gamma^{\rho}\partial/\partial k_{\rho}$. However, a direct result from such
an operation is the appearance of a higher power denominator, $1/\left[(p_V+k_l)^2-m^2\right]^2$,
when the extra derivative operates on the momentum $k$. Thus we have to use the following trick 
\begin{equation}
{\rm Disc}_{p_V^2}\left[\int \frac{1}{\left[(p_V+k_l)^2-m^2\right]^2}(\cdots)\right]=(-2\pi i)
\frac{\partial}{\partial\Omega}\int \text{\ \ensuremath{\delta}}\left[
(p_V+k_l)^2-\Omega\right](\cdots)\Big|_{\Omega=m^{2}}~,
\end{equation}
to extract the discontinuity. Note that the auxiliary parameter $\Omega$ should be set to $m$ at
the end of all the integrations. The imaginary part of the $F_{V(i)}^{{\cal O}_1}$  from the $\slashed x$
term of Eq.~(\ref{BLCDAs}) then is
\begin{align}
&{\rm Im}^2F_{V(i)}^{{\cal O}_1}(p_{\pi}^2, p_{V}^2)_{(a)\slashed x}  \nonumber\\
=&\frac{-i \pi^5 f_B m_B}{\sqrt{2}}\int_0^{\infty}d\omega\left(\tilde\phi_{-}^{B}(\omega)-\tilde
\phi_{+}^{B}(\omega)\right)\delta\left[\left(1-\frac{\omega}{m_B}\right)p_{\pi}^2-\omega m_B+\omega^2-m^2
\right]\nonumber\\
&\times\Big\{\int d\Phi_2(p_V^2,k_2,k_3)\ {\cal N}^{V(i)}_{(a)}(p_{\pi},k_2,k_3)+\frac{\partial}
{\partial\Omega}\int d\Phi_2(p_V^2,k_2,k_3)_{\Omega}\ {\cal L}^{V(i)}_{(a)}(p_{\pi},k_2,k_3)\Big\}
\Big|_{\Omega=m^{2}},\label{Correx0ImpV}
\end{align}
where 
\begin{equation}
\int d\Phi_2(p_V^2,k_2,k_3)_{\Omega}=\int \frac{d^3 \vec{k}_2}{(2\pi)^3}\frac{1}{2E_{k_2}}
\frac{d^3 \vec{k}_3}{(2\pi)^3}\frac{1}{2E_{k_3}}\delta^{(4)}(p_V-k_2-k_3)\delta(k_2^2-m^2)\delta(k_3^2-\Omega).
\end{equation}
The calculation for the contribution from diagram (b) in Fig.~\ref{fig:FeynmanDiag} is similar except
that the loop bubble is at the corner of pion the vertex. Therefore, now the imaginary part in terms of
$p_{\pi}^2$ comes from the loop integration, while the imaginary part in terms of $p_{V}^2$ comes
from the rest of the  propagator. The imaginary part from the diagram (b) follows as
\begin{align}
&{\rm Im}^2F_{V(i)}^{{\cal O}_1}(p_{\pi}^2, p_{V}^2)_{(b)} \nonumber\\
=&\  3 i \sqrt{2} \pi^5 f_B m_B\int_0^{\infty}d\omega\ \phi_{+}^{B}(\omega)\delta\left[\left(1
-\frac{\omega}{m_B}\right)p_{V}^2-\omega m_B+\frac{\omega}{m_B}m_{\pi}^2+\omega^2-m^2\right] \nonumber\\
&\times\int d\Phi_2(p_V^2,k_2,k_3)\ {\cal M}^{V(i)}_{(b)}(p_{\pi},k_2,k_3)\nonumber\\
&+\frac{3 i \pi^5 f_B m_B}{\sqrt{2}}\int_0^{\infty}d\omega\left(\tilde\phi_{-}^{B}(\omega)-\tilde
\phi_{+}^{B}(\omega)\right)\delta\left[\left(1-\frac{\omega}{m_B}\right)p_{V}^2-\omega m_B+\frac{\omega}{m_B}
m_{\pi}^2+\omega^2-m^2\right]\nonumber\\
&\times\int d\Phi_2(p_V^2,k_2,k_3)\ {\cal N}^{V(i)}_{(b)}(p_{\pi},k_2,k_3).\nonumber\\
&+\frac{3 i \pi^5 f_B m_B}{\sqrt{2}}\int_0^{\infty}d\omega\left(\tilde\phi_{-}^{B}(\omega)
-\tilde\phi_{+}^{B}(\omega)\right)\frac{\partial}{\partial\Omega}\left\{\delta\left[\left(1
  -\frac{\omega}{m_B}\right)p_{V}^2-\omega m_B+\frac{\omega}{m_B}m_{\pi}^2+\omega^2-\Omega\right]\right.
\nonumber\\
&\left.\times\int d\Phi_2(p_V^2,k_2,k_3)_{\Omega}\ {\cal L}^{V(i)}_{(b)}(p_{\pi},k_2,k_3)\right\}
\Big|_{\Omega=m^{2}}.
\label{Correx0ImpV}
\end{align}
For the cases of scalar and tensor resonances, the imaginary part of the corresponding correlation
function can be derived similarly. For simplicity,  we only present the results of the
corresponding calculations. For the scalar resonance, only the diagram (b) contributes in the
approximation $m=0$, and the expression of ${\rm Im}^2 F_{S}^{{\cal O}_1}$ reads 
\begin{align}
&{\rm Im}^2F_{S}^{{\cal O}_1}(p_{\pi}^2, p_S^2) \nonumber\\
=&\  3\sqrt{2}i\ {\rm sin}\alpha\   \pi^5 f_B m_B\int_0^{\infty}d\omega\ \phi_{+}^{B}(\omega)\delta\left[
\left(1-\frac{\omega}{m_B}\right)p_T^2-\omega m_B+\frac{\omega}{m_B}m_{\pi}^2+\omega^2-m^2\right]
\nonumber\\&\times\int d\Phi_2(p_S^2,k_2,k_3)\ {\cal M}^{S}_{(b)}(p_{\pi},k_2,k_3)\nonumber\\
&+\frac{3 i\ {\rm sin}\alpha\   \pi^5 f_B m_B}{\sqrt{2}}\int_0^{\infty}d\omega\left(\tilde\phi_{-}^{B}(\omega)
-\tilde\phi_{+}^{B}(\omega)\right)\delta\left[\left(1-\frac{\omega}{m_B}\right)p_S^2-\omega m_B
+\frac{\omega}{m_B}m_{\pi}^2+\omega^2-m^2\right]\nonumber\\
&\times\int d\Phi_2(p_S^2,k_2,k_3)\ {\cal N}^{S}_{(b)}(p_{\pi},k_2,k_3).\nonumber\\
&+\frac{3 i\ {\rm sin}\alpha\   \pi^5 f_B m_B}{\sqrt{2}}\int_0^{\infty}d\omega\left(\tilde\phi_{-}^{B}(\omega)
-\tilde\phi_{+}^{B}(\omega)\right)\frac{\partial}{\partial\Omega}\left\{\delta\left[\left(1-
\frac{\omega}{m_B}\right)p_S^2-\omega m_B+\frac{\omega}{m_B}m_{\pi}^2+\omega^2-\Omega\right]\right.\nonumber\\
&\left.\times\int d\Phi_2(p_T^2,k_2,k_3)_{\Omega}\ {\cal L}^{S}_{(b)}(p_{\pi},k_2,k_3)\right\} \Big|_{\Omega=m^{2}}.
\label{Correx0ImpS}
\end{align}
For the tensor resonance, like the case of vector resonance, both the diagram (a) and (b) contribute to
the correlation function. The expression of ${\rm Im}^2 F_{T(i)}^{{\cal O}_1}$ is given by
\begin{align}
&{\rm Im}^2F_{T(i)}^{{\cal O}_1}(p_{\pi}^2, p_T^2)={\rm Im}^2F_{T(i)}^{{\cal O}_1}(p_{\pi}^2, p_T^2)_{(a)}+{\rm Im}^2F_{T(i)}^{{\cal O}_1}(p_{\pi}^2, p_T^2)_{(b)},\\
&{\rm Im}^2F_{T(i)}^{{\cal O}_1}(p_{\pi}^2, p_T^2)_{(a)}\nonumber\\
=&\frac{-i \pi^5 f_B m_B}{\sqrt{2}}\int_0^{\infty}d\omega\ \phi_{+}^{B}(\omega)\delta\left[\left(1-\frac{\omega}{m_B}\right)p_{\pi}^2-\omega m_B+\omega^2-m^2\right] \nonumber\\
&\times\int d\Phi_2(p_T^2,k_2,k_3)\ {\cal M}^{T(i)}_{(a)}(p_{\pi},k_2,k_3)\nonumber\\
&-\frac{ i \pi^5 f_B m_B}{2\sqrt{2}}\int_0^{\infty}d\omega\left(\tilde\phi_{-}^{B}(\omega)-\tilde\phi_{+}^{B}(\omega)\right)\delta\left[\left(1-\frac{\omega}{m_B}\right)p_{\pi}^2-\omega m_B+\omega^2-m^2\right] \nonumber\\
&\times\int d\Phi_2(p_T^2,k_2,k_3)\ {\cal N}^{T(i)}_{(a)}(p_{\pi},k_2,k_3).\nonumber\\
&-\frac{ i \pi^5 f_B m_B}{2\sqrt{2}}\int_0^{\infty}d\omega\left(\tilde\phi_{-}^{B}(\omega)-\tilde\phi_{+}^{B}(\omega)\right)\delta\left[\left(1-\frac{\omega}{m_B}\right)p_{\pi}^2-\omega m_B+\omega^2-m^2\right]\nonumber\\
&\times\frac{\partial}{\partial\Omega}\left\{\int d\Phi_2(p_T^2,k_2,k_3)_{\Omega}\ {\cal L}^{T(i)}_{(a)}(p_{\pi},k_2,k_3)\right\} \Big|_{\Omega=m^{2}},\label{Correx0ImpTa}\\
&{\rm Im}^2F_{T(i)}^{{\cal O}_1}(p_{\pi}^2, p_T^2)_{(b)} \nonumber\\
=&\  \frac{3 i  \pi^5 f_B m_B}{\sqrt{2}}\int_0^{\infty}d\omega\ \phi_{+}^{B}(\omega)\delta\left[\left(1-\frac{\omega}{m_B}\right)p_T^2-\omega m_B+\frac{\omega}{m_B}m_{\pi}^2+\omega^2-m^2\right] \nonumber\\&\times\int d\Phi_2(p_T^2,k_2,k_3)\ {\cal M}^{T(i)}_{(b)}(p_{\pi},k_2,k_3)\nonumber\\
&+\frac{3 i  \pi^5 f_B m_B}{2\sqrt{2}}\int_0^{\infty}d\omega\left(\tilde\phi_{-}^{B}(\omega)-\tilde\phi_{+}^{B}(\omega)\right)\delta\left[\left(1-\frac{\omega}{m_B}\right)p_T^2-\omega m_B+\frac{\omega}{m_B}m_{\pi}^2+\omega^2-m^2\right]\nonumber\\
&\times\int d\Phi_2(p_T^2,k_2,k_3)\ {\cal N}^{T(i)}_{(b)}(p_{\pi},k_2,k_3).\nonumber\\
&+\frac{3 i  \pi^5 f_B m_B}{2\sqrt{2}}\int_0^{\infty}d\omega\left(\tilde\phi_{-}^{B}(\omega)-\tilde\phi_{+}^{B}(\omega)\right)\frac{\partial}{\partial\Omega}\left\{\delta\left[\left(1-\frac{\omega}{m_B}\right)p_T^2-\omega m_B+\frac{\omega}{m_B}m_{\pi}^2+\omega^2-\Omega\right]\right.\nonumber\\
&\left.\times\int d\Phi_2(p_T^2,k_2,k_3)_{\Omega}\ {\cal L}^{T(i)}_{(b)}(p_{\pi},k_2,k_3)\right\} \Big|_{\Omega=m^{2}}.
\label{Correx0ImpTb}
\end{align}
The explicit expressions for the ${\cal M}, {\cal N}$ and ${\cal L}$ functions are given in
Appendix~C. The LCSR
calculation for the ${\cal O}_{2\cdots 7}$ matrix elements can be done similarly, so we will not
explicitly present them here.

\section{Phenomenological Results}\label{sec:phenomenology}

\subsection{Numerical Results for $\langle R, M\left | {\cal O}_i\right|B^-\rangle$}
First, we list the values  of all the parameters used in this work. All the mass parameters are:
$m_u=m_d=m=0$, $m_s=93$~MeV($\mu=2$ GeV), 
$m_{\pi}=0.139$~GeV, $m_{K^{\pm}}=0.496$~GeV, $m_B=5.28$~GeV,
$m_{f_0(980)}=0.99$~GeV, $m_{\rho(770)}=0.775$~GeV, $m_{\rho(1450)}=1.465$~GeV, $m_{f_2(1270)}=1.275$~GeV,
$m_{\phi(1020)}=1.02$~GeV, $m_{K_0^*(1430)}=1.43$~GeV, $m_{K^*(892)}=0.892$~GeV, $m_{f_0(1370)}=1.37$~GeV,
$m_{\rho(1570)}=1.57$~GeV,  $m_{f_2^{\prime}(1525)}=1.52$~GeV, $m_{\phi(1680)}=1.68$~GeV, $m_{K^*(1410)}=1.41$~GeV,
$m_{K_0(1950)}=1.95$~GeV \cite{Zyla:2020zbs}.

The value of all the decay constants used here are: $f_B=0.207$~GeV \cite{Gelhausen:2013wia},
$f_{\pi}=0.13$~GeV \cite{Gasser:2010wz}, $f_{\rho(770)}=0.21$~GeV \cite{Chang:2018aut},
$f_{\rho(1450)}=0.186$~GeV \cite{Li:2017mao}, $f_{f_2(1270)}=0.102$~GeV \cite{Cheng:2010yd},
$f_{\phi(1020)}=0.102$~GeV \cite{Chen:2020qma}, $f_{K}=0.11$~GeV \cite{Guo:2018pyq},
$f_{K^*(892)}=0.204$~GeV \cite{Chang:2018aut}, $f_{K_0^*(1430)}=0.427$~GeV  \cite{Du:2004ki}.

For the leading twist LCDAs of the $B$ meson, their expressions as well as the associated
coefficients are taken from  \cite{Braun:2017liq}
\begin{eqnarray}
\phi_B^{+}(\omega) &=& {\omega \over \omega_0^2} \,  e^{-\omega/\omega_0} \,, \nonumber \\
\phi_B^{-}(\omega) &=& {1 \over \omega_0} \,  e^{-\omega/\omega_0}
- {\lambda_E^2 - \lambda_H^2 \over 9 \, \omega_0^3}  \,
\left [ 1 - 2 \, \left ( {\omega \over \omega_0} \right )
+ {1 \over 2} \, \left ( {\omega \over \omega_0} \right )^2 \right ]
\,  e^{-\omega/\omega_0}~,
\end{eqnarray}
where $\omega_0=(2/3)\bar \Lambda$, $ \lambda_H^2=2 \lambda_E^2$ and $\lambda_H=\bar \Lambda$, with
$\bar \Lambda=m_B-m_b=0.45$~GeV in the heavy quark limit \cite{Ball:1993xv}.
The numerical result for the $T_{{\cal O}_{i}}^{R}$ are listed in Table~\ref{tab:NumericalMat}, where
$s_R$ is the threshold parameter for the resonance, while $s_{\pi}$ and $s_K$ are the threshold parameters
for the non-resonant particle, namely the final pion or kaon. The Borel parameters are chosen in
the region where the numerical values of $T_{{\cal O}_{i}}^{R}$ are stable. The errors of the
$T_{{\cal O}_{i}}^{R}$  come from this uncertainties of the Borel parameters.
\begin{table}
\caption{Numerical results of the parameters for the matrix element $\langle R, M\left | {\cal O}_i\right
|B^-\rangle$, where the threshold parameter for the final pion or kaon   are $s_{\pi}=m_{\pi(1300)}^2$ or
$s_K=m_{K(1460)}^2$ respectively .}
\label{tab:NumericalMat}
\begin{tabular}{|c|ccccc|}
\hline 
\hline 
$T_{{\cal O}_{i}}^{R}$ & Value & $~~~~s_{R}~~~$ & $~~s_{\pi}$ or $s_{K}~~$ & ~$M_{1}$ (GeV)~~ & ~~$M_{2}$ (GeV)~~\tabularnewline
\hline 
\hline 
$T_{{\cal O}_{1}}^{f_{0}(980)}$ & $-0.23\pm0.05$ ${\rm GeV}^{3}$ & $m_{f_{0}(1370)}^{2}$ & $s_{\pi}$ & $10\pm1$ & $0.4\pm0.05$\tabularnewline
$T_{{\cal O}_{1}}^{\rho(770)}$ & $0.043\pm0.013$ ${\rm GeV}^{2}$ & $m_{\rho(1450)}^{2}$ & $s_{\pi}$ & $5\pm1$ & $0.15\pm0.05$\tabularnewline
$T_{{\cal O}_{2}}^{\rho(770)}$ & $0.015\pm0.003$ ${\rm GeV}^{2}$ & $m_{\rho(1450)}^{2}$ & $s_{\pi}$ & $6\pm1$ & $0.15\pm0.05$\tabularnewline
$T_{{\cal O}_{3}}^{\rho(770)}$ & $(3\pm1)\times10^{-4}$ ${\rm GeV}^{2}$ & $m_{\rho(1450)}^{2}$ & $s_{\pi}$ & $6\pm1$ & $0.15\pm0.05$\tabularnewline
$T_{{\cal O}_{1}}^{\rho(1450)}$ & $0.014\pm0.008$ ${\rm GeV}^{2}$ & $m_{\rho(1570)}^{2}$ & $s_{\pi}$ & $10\pm1$ & $0.2\pm0.05$\tabularnewline
$T_{{\cal O}_{1}}^{f_{2}(1270)}$ & $0.0021\pm0.0006$ ${\rm GeV}$ & $m_{f_{2}^{\prime}(1525)}$ & $s_{\pi}$ & $10\pm1$ & $0.15\pm0.05$\tabularnewline
$T_{{\cal O}_{2}}^{\rho(1450)}$ & $0.0055\pm0.002$ ${\rm GeV}^{2}$ & $m_{\rho(1570)}^{2}$ & $s_{\pi}$ & $8\pm1$ & $0.2\pm0.05$\tabularnewline
$T_{{\cal O}_{3}}^{\rho(1450)}$ & $(1.3\pm1.0)\times10^{-4}$ ${\rm GeV}^{2}$ & $m_{\rho(1570)}^{2}$ & $s_{\pi}$ & $10\pm1$ & $0.1\pm0.05$\tabularnewline
$T_{{\cal O}_{4}}^{\phi(1020)}$ & $0.003\pm0.0016$ ${\rm GeV}^{2}$ & $m_{\phi(1680)}^{2}$ & $s_{\pi}$ & $8\pm1$ & $0.3\pm0.05$\tabularnewline
$T_{{\cal O}_{5}}^{f_{0}(980)}$ & $(1.5\pm1.0)\times10^{-4}$ ${\rm GeV}^{3}$ & $m_{f_{0}(1370)}^{2}$ & $s_{\pi}$ & $5\pm1$ & $0.1\pm0.05$\tabularnewline
$T_{{\cal O}_{6}}^{K^{*}(892)}$ & $0.176\text{\ensuremath{\pm}0.009}$ ${\rm GeV}^{2}$ & $m_{K^{*}(1410)}^{2}$ & $s_{K}$ & $8\pm1$ & $3\pm1$\tabularnewline
$T_{{\cal O}_{6}}^{K_{0}(700)}$ & $-0.093\pm0.004$ ${\rm GeV}^{3}$ & $m_{K_{0}^*(1430)}^{2}$ & $s_{K}$ & $8\pm1$ & $0.36\pm0.05$\tabularnewline
$T_{{\cal O}_{7}}^{K_{0}(700)}$ & $0.222\pm0.02$ ${\rm GeV}^{3}$ & $m_{K_{0}^*(1430)}^{2}$ & $s_{K}$ & $6\pm1$ & $0.4\pm0.05$\tabularnewline
$T_{{\cal O}_{6}}^{K_{0}^*(1430)}$ & $-0.018\pm0.004$ ${\rm GeV}^{3}$ & $m_{K_{0}(1950)}^{2}$ & $s_{K}$ & $8\pm1$ & $0.4\pm0.05$\tabularnewline
$T_{{\cal O}_{7}}^{K_{0}(1430)}$ & $0.14\pm0.017$ ${\rm GeV}^{3}$ & $m_{K_{0}(1950)}^{2}$ & $s_{K}$ & $6\pm1$ & $0.45\pm0.05$\tabularnewline
\hline 
\end{tabular}
\end{table}

\subsection{Branching fractions for the $B^-\to K^+K^-\pi^-$ decay }

In a first step, we give a general formula for the calculation of three-body decays. We define
three dimensionless Lorentz invariant variables:
\begin{equation}
x_1=\frac{2 p_B\cdot p_{\pi^-}}{m_B^2},~~~x_2=\frac{2 p_B\cdot p_{K^+}}{m_B^2},~~~
x_3=\frac{2 p_B\cdot p_{K^-}}{m_B^2}.
\end{equation}
Thus, the three Mandelstam variables can be expressed by the $x_i$ as
\begin{align}
&s=(p_{K^+}+p_{K^-})^2=m_B^2(1+\lambda_{\pi}-x_1),\nonumber\\
&t=(p_{\pi^-}+p_{K^+})^2=m_B^2(1+\lambda_{K}-x_3),\nonumber\\
&u=(p_{\pi^-}+p_{K^-})^2=m_B^2(1+\lambda_{K}-x_2),
\end{align}
where $\lambda_{K}=m_{K}^2/m_B^2$ and the $x_i$ satisfy $x_1+x_2+x_3=2$. Each Lorentz invariant
quantity in the decay amplitude can be expressed by the $x_i$. In the rest-frame of the $B$ meson,
the formula for the three-body decay width is
\begin{equation}
\Gamma\left[B^-\to K^+K^-\pi^-\right]=\frac{m_B}{256\pi^3}\int_S dx_2 dx_3\ {\cal A}(x_1,x_2,x_3),
\end{equation}
where ${\cal A}(x_1,x_2,x_3)$ is the total decay amplitude. The integration region $S$ is:
\begin{equation}
-1<\frac{ (x_2 x_3-2 x_2-2 x_3+2)+4
   \lambda_K-2 \lambda_{\pi}}{\sqrt{
   x_2^2-4 \lambda_K} \sqrt{ x_3^2-4
   \lambda_{K}}}<1, \label{x2x3region}
\end{equation}
as well as $2\sqrt{\lambda_K}<x_2<2(1-\sqrt{\lambda_K}-\sqrt{\lambda_{\pi}})$ and $2\sqrt{\lambda_K}
<x_3<2(1-\sqrt{\lambda_{\pi}})-x_2$.
The boundary values $\pm 1$ in Eq.~(\ref{x2x3region}) are reached when the direction of
the $K^+$ three-momentum is parallel or anti-parallel with that of the $K^-$, 

All the strong coupling constants are taken from the Eq.~(2.40) of Ref.~\cite{Cheng:2020ipp}.
The decays widths of the resonances are  \cite{Zyla:2020zbs}: 
\begin{align}
&\Gamma_{K_0(1430)}=0.27~{\rm GeV},~~~\Gamma_{K^*(892)}=0.051~{\rm GeV},~~~\Gamma_{\rho(1450)}=0.4~{\rm GeV},
\nonumber\\
&\Gamma_{f_2(1270)}=0.187~{\rm GeV},~~~\Gamma_{f_0(980)}=0.05~{\rm GeV},~~~\Gamma_{\phi(1020)}=0.0042~{\rm GeV}.
\end{align}
\begin{table}
\caption{Decay branching fraction from the various resonances (in units of $10^{-6}$). We also compare
our results with those from experiments $\cal{B}_{\rm expt}$ and those from Ref.\cite{Cheng:2020ipp}.
The $\cal{B}_{\rm expt}$ are inferred from the measured branching fraction of each resonance \cite{LHCb:2019xmb} and the world averaged branching fraction of the entire decay process $\mathcal{B}_{\rm averaged}
\left(B^{\pm}\rightarrow \pi^{\pm} K^{+} K^{-}\right)=(5.24 \pm 0.42) \times 10^{-6}$ \cite{HFLAV:2019otj}.}
\label{ResonaBranch}
\begin{tabular}{|c|ccc|}
\hline 
\hline 
Contribution & ${\cal B}_{{\rm expt}}$ & Ref.\cite{Cheng:2020ipp} & This work\tabularnewline
\hline 
\hline 
$K^{*}(890)$ & $0.39\pm0.05$ & $0.23_{-0.04}^{+0.04}$ & $0.014\pm0.0015$\tabularnewline
$K_{0}^{*}(1430)$ & $0.23\pm0.08$ & $0.71_{-0.12}^{+0.13}$ & $0.11\pm0.025$\tabularnewline
$\rho(1450)$ & $1.61\pm0.15$ & \textendash{} & $0.035\pm0.04$\tabularnewline
$f_{2}(1270)$ & $0.39\pm0.06$ & $0.05_{-0.01}^{+0.01}$ & $0.44\pm0.24$\tabularnewline
$\phi(1020)$ & $0.016\pm0.008$ & $0.0079_{-0.0017}^{+0.0019}$ & $(0.97\pm0.76)\times10^{-5}$\tabularnewline
$f_{0}(980)$ & \textendash{} & $0.19_{-0.03}^{+0.03}$ & $0.12\pm0.04$\tabularnewline
\hline 
\end{tabular}
\end{table}
The decay branching fraction stemming from the various resonances are listed in Table~\ref{ResonaBranch}.
We also compare our results with those from experiments $\cal{B}_{\rm expt}$ and those from
Ref~\cite{Cheng:2020ipp}. The $\cal{B}_{\rm expt}$  are inferred from the measured branching fraction
of each resonance \cite{LHCb:2019xmb} as well as the branching fraction of the entire decay
process $\mathcal{B}\left(B^{\pm} \rightarrow \pi^{\pm} K^{+} K^{-}\right)=(5.24 \pm 0.42) \times 10^{-6}$
\cite{HFLAV:2019otj}. We note that the $f_2(1270)$ and $K^*_0(1430)$ contributions obtained in this work is consistent
with those from experiment. Our result for the $f_0(980)$ is consistent with that from
Ref.~\cite{Cheng:2020ipp}, which is still waiting for future experimental tests. The $\phi(1020)$ contribution is much smaller than the other
determinations, which is due to the tiny value of the Wilson coefficients $a_3, a_5, a_7$ and $a_9$
as shown in Eq.~(\ref{WilsonCoeff}). The other fractions are about one order smaller than the
experimental values. A possible reason for this difference is due to the uncertainty of the
strong couplings, which require further precise measurements in the future. The contribution
of all the resonances considered above to the branching fraction is
\begin{equation}
{\cal B}_{\rm resonant}\left[B^-\to K^+K^-\pi^-\right]=(0.54\pm 0.23)\times 10^{-6}.
\end{equation}

Thus,  the resonant contribution obtained above is only a small part of the total branching
fraction $\mathcal{B}\left(B^{\pm} \rightarrow \pi^{\pm} K^{+} K^{-}\right)=(5.24 \pm 0.42) \times 10^{-6}$
\cite{HFLAV:2019otj}. Besides the resonances, the non-resonant (NR) contributions also contribute to
the decay process.  In our case, it comes from the matrix element  of ${\cal O}_1$ and ${\cal O}_7$.
The ${\cal O}_1$ matrix element can be generally parameterized by
\begin{align}
&\langle\pi^{-}(p_{1})K^{+}(p_{2})K^{-}(p_{3})|{\cal O}_1|B^{-}(p)\rangle_{NR}\nonumber\\
=&\ -\frac{f_{\pi}}{2}\left[2 m_{\pi}^{2} r+\left(m_{B}^{2}-s_{23}-m_{\pi}^{2}\right) \omega_{+}+\left(s_{12}
-s_{13}\right) \omega_{-}\right].
\end{align}
The form factors $r, \omega_{\pm}$ are calculated within heavy meson chiral perturbation theory (HMChPT)
\cite{Lee:1992ih}. However, since HMChPT is only reliable in the low-energy region, in the case of
$B$ decays, the final-state energy is high enough so that a direct use of HMChPT will make the amplitude
blow up. There is a practical method to solve this problem. As proposed by Ref.~\cite{Cheng:2007si},
one can introduce an exponential term to suppress the amplitude at high energy, and thus
the ${\cal O}_1$ matrix element is modified as
\begin{align}
\langle\pi^{-}(p_{1})K^{+}(p_{2})K^{-}(p_{3})|{\cal O}_1|B^{-}(p)\rangle=&\ \langle\pi^{-}(p_{1})K^{+}(p_{2})
K^{-}(p_{3})|{\cal O}_1|B^{-}(p)\rangle_R\nonumber\\ +&\ e^{-\alpha_{\rm NR}p_B\cdot(p_{2}+p_{3})}\langle\pi^{-}(p_{1})
K^{+}(p_{2})K^{-}(p_{3})|{\cal O}_1|B^{-}(p)\rangle_{NR},
\end{align}
where $\alpha_{NR}=0.16$~GeV$^{-2}$\cite{Cheng:2020ipp}. On the other hand, we will not include the
NR contribution of ${\cal O}_7$. The first reason is that as shown by the Eqs.~(2.22)-(2.24) of
Ref.~\cite{Cheng:2007si}, such NR contribution depends on the NR component of a scalar matrix
element $\left\langle\pi^{-}\left(p_{1}\right) K^{+}\left(p_{2}\right)|\bar{d} s| 0\right\rangle^{\mathrm{NR}}$.
This scalar matrix element itself has been obtained in Ref.~\cite{Shi:2020rkz}. However, its
exact NR component is still unknown. Although Refs.~\cite{Cheng:2013dua,Cheng:2016shb} proposed a
formula to characterize its NR component, there still exists an unknown phase parameter.
Another reason is that the Wilson coefficients $a_6^p$ and $a_8^p$ corresponding to ${\cal O}_7$
are suppressed compared with that of ${\cal O}_1$. Thus, in practice we  only consider the NR
effect in the ${\cal O}_1$ matrix element.

The final-state interactiom (FSI) of $\pi^{+} \pi^{-} \leftrightarrow K^{+} K^{-}$ also contributes
to the decay branching fraction. The rescattering amplitude reads \cite{Cheng:2020ipp}
\begin{align}
{\cal A}\left(B^{-} \rightarrow K^{+} K^{-} \pi^{-}\right)_{\text {FRS}} &=e^{i \delta_{\pi \pi}}\left[\cos
(\phi / 2) {\cal A}\left(B^{-} \rightarrow (f_0(980)\to K^{+} K^{-}) \pi^{-}\right)\right.\nonumber\\
&\left.+i \sin (\phi / 2) {\cal A}\left(B^{-} \rightarrow (\sigma(500)\to\pi^{+} \pi^{-}) \pi^{-}\right)
\right]
\end{align}
where the re-scattering mixes the S-wave components of $B^{-} \rightarrow K^{+} K^{-} \pi^{-}$ and $B^{-}
\rightarrow \pi^{+} \pi^{-} \pi^{-}$. The mixture coefficients $\delta_{\pi \pi}$ and $\phi$ are functions
of the S-wave invariant mass square $s_{23}$ which are given in Ref.~\cite{Cheng:2016shb}. The
calculation of $\langle \sigma(500)(p-p_{1})\pi^{-}(p_{1})|{\cal O}_{1,5}|B^{-}(p)\rangle$ is similar to
that of $f_0(980)$, and the results are
\begin{equation}
T_{{\cal O}_1}^{\sigma(500)}=-0.033\pm0.008~{\rm GeV}^3,~~~T_{{\cal O}_5}^{\sigma(500)}=(1.2\pm0.08)
\times 10^{-4}~{\rm GeV}^3.
\end{equation}
The mass and decay constant used here is $m_{\sigma(500)}=0.5$~GeV\cite{Zyla:2020zbs} and
$f_{\sigma(500)}=0.89$~GeV\cite{Qi:2018wkj}. The decay width of the $\sigma(500)$ is $\Gamma_{\sigma(500)}
=0.3$~GeV\cite{Zyla:2020zbs}.  Finally, combining the resonant, NR and FSI contributions, we obtain
the total decay branching fraction:
\begin{equation}
{\cal B}_{\rm total}\left[B^-\to K^+K^-\pi^-\right]=(0.87\pm 0.24)\times 10^{-6},
\end{equation}
which is of the same order as the world averaged value, but is still sizeably smaller than that.
The possible reasons for this discrepancy are:
\begin{itemize}
\item In this work, we do not consider the contributions from the higher twist LCDAs of the $B$ meson
  since they are expected to be one order suppressed according to our previous work \cite{Shi:2021bvy}.
  However, to remedy this part, in the future we will perform a more in-depth study on the
  contribution of higher twist LCDAs.
\item The NR contribution to the ${\cal O}_7$ matrix element is not introduced to the total branching
  fraction. We have argued that it is suppressed by its Wilson coefficients $a_6^p, a_8^p$. However,
  in principle its exact contribution should also depend on the model parameters introduced in
  Ref.~\cite{Cheng:2007si}, but they are not determined unambiguously.
\item Generally, the matrix elements $\langle R, M\left | {\cal O}_i\right|B^-\rangle$ contains a
  strong phase which cannot be obtained by the approach used in this work. The interference between
  these strong phases will affect the amount of the total branching fraction.  In principle, such
  strong phase can be produced by the charming or strange penguin loop which was studied by
  Ref.~\cite{Khodjamirian:2003eq} for the case of $B\to \pi\pi$ decay. Similarly, for the $B \to M+ R$
  decay, the penguin loop can also produce a strong phase, which is expected to be studied in the future.
\end{itemize}

\section{Conclusion}\label{sec:conclusion}
In this work, we have studied the resonant contribution to the decay amplitude of $B^-\to K^+K^-\pi^-$,
which is dominated by the scalar $f_0(980), K_0^*(1430)$, vector $\rho(1450), \phi(1020), K^*(892)$ and
tensor $f_2(1270)$ resonances. The three-body decay is reduced to various quasi two-body
decays, where the $B$ meson first decays into a resonance and a pion or a kaon, and subsequently the
resonance decays into the other two final-state mesons. The quasi two-body decays are calculated within
the LCSR approach using the leading twist $B$ meson LCDAs. Then, we calculated the decay branching
fraction for $B^-\to K^+K^-\pi^-$ from each of the resonances considered here. Some of them are consistent
with experiment while the others are smaller than the measured values. One possible reason for this
discrepancy are the uncertainties of the strong couplings between the corresponding resonance with
pseudoscalar mesons. Including the effects of the non-resonant and final state rescattering effects,
we also evaluate the total branching fraction and the result is smaller than the world-averaged value. We have listed three possible reasons for this discrepancy, which requires further researchs in the future.

\section*{Acknowledgements}
This work is supported in part by the NSFC and the Deutsche Forschungsgemeinschaft (DFG, German Research
Foundation) through the funds provided to the Sino-German Collaborative
Research Center TRR110 ``Symmetries and the Emergence of Structure in QCD''
(NSFC Grant  No. 12070131001, DFG Project-ID 196253076 - TRR 110). 
The work of UGM was supported in part by the Chinese
Academy of Sciences (CAS) President's International
Fellowship Initiative (PIFI) (Grant No. 2018DM0034)
and by VolkswagenStiftung (Grant No. 93562).

\begin{appendix}

\section{ Expression of $\xi_i(\lambda_{\pi},\lambda_T)$}

The expressions of $\xi_i(\lambda_{\pi},\lambda_T)$ given in the end of Section \ref{sec:lc_sum_rules} are
\begin{align}
\xi_1(\lambda_{\pi},\lambda_T) =&-\frac{1}{\lambda_T}\left[\lambda_{\pi}^4+2 \lambda_{\pi}^3(\lambda_{T}-2)+\lambda_{\pi}^2 \left(4 \lambda_{T}^2-8 \lambda_{T}+6\right)+2 \lambda_{\pi}
   (\lambda_{T}-2)(\lambda_{T}-1)^2+(\lambda_{T}-1)^4\right],\nonumber\\
\xi_2(\lambda_{\pi},\lambda_T) =&\frac{1}{2 (\lambda_{\pi}+\lambda_{T}-1)^3} \left[\lambda_{\pi}^6+2 \lambda_{\pi}^5
   (\lambda_{T}-3)+\lambda_{\pi}^4
   \left(5 \lambda_{T}^2-14
   \lambda_{T}+15\right)\right.\nonumber\\
   &\left.+4
   \lambda_{\pi}^3
   \left(\lambda_{T}^3-6
   \lambda_{T}^2+9
   \lambda_{T}-5\right)+\lambda_{\pi}^2 (\lambda_{T}-1)^2
   \left(5 \lambda_{T}^2-14
   \lambda_{T}+15\right)\right.\nonumber\\
   &\left.+2
   \lambda_{\pi}
   (\lambda_{T}-3)
   (\lambda_{T}-1)^4+(\lambda_{T}-1)^6\right],\nonumber\\
\xi_3(\lambda_{\pi},\lambda_T) =& - \left[\lambda_{\pi}^6+2
   \lambda_{\pi}^5
   (\lambda_{T}-3)+\lambda_{\pi}^4
   \left(5 \lambda_{T}^2-14
   \lambda_{T}+15\right)+4
   \lambda_{\pi}^3
   \left(\lambda_{T}^3-6
   \lambda_{T}^2+9
   \lambda_{T}-5\right)\right.\nonumber\\
   &\left.+\lambda_{\pi}^2 (\lambda_{T}-1)^2
   \left(5 \lambda_{T}^2-14
   \lambda_{T}+15\right)+2
   \lambda_{\pi}
   (\lambda_{T}-3)
   (\lambda_{T}-1)^4+(\lambda_{T}-1)^6\right]\Big / \nonumber\\
   &\left[\lambda_{\pi}^4+4 \lambda_{\pi}^3
   (\lambda_{T}-1)+2
   \lambda_{\pi}^2 \left(5
   \lambda_{T}^2-6
   \lambda_{T}+3\right)+4
   \lambda_{\pi}
   (\lambda_{T}-1)^3+(\lambda_{T}-1)^4\right].
\end{align}

\section{Two-body phase space integration}
In this appendix we give  the tensor bases used in this work for the two-body phase space integration. The rank-0 integration is defined as 
\begin{align}
\Phi_2^{(0)}(s,m_1,m_2)\equiv&\int \frac{d^3 \vec{k}_1}{(2\pi)^3}\frac{1}{2E_{k_2}}\frac{d^3 \vec{k}_2}{(2\pi)^3}\frac{1}{2E_{k_2}}\delta^{(4)}(p-k_1-k_2)\delta(k_1^2-m^1)\delta(k_2^2-m_2^2)\nonumber\\
\equiv&\int d\Phi_2=\ \frac{\pi\sqrt{\left(s-(m_{1}+m_{2})^{2}\right)\left(s-(m_{1}-m_{2})^{2}\right)}}{(2\pi)^{6}2s}
\end{align}
with $s=p^2$. The higher rank integrations are defined as
\begin{align}
\int d\phi_2 \ k_1^{\mu}&=\Phi_2^{(0)}(s,m_1,m_2) A_1 p^{\mu},\\
\int d\phi_2 \ k_1^{\mu}k_2^{\nu}&=\Phi_2^{(0)}(s,m_1,m_2) \left[A_2 p^{\mu}p^{\nu}+B_2 p^2 g^{\mu\nu}\right],\\
\int d\phi_2 \ k_1^{\mu}k_2^{\nu}k_2^{\rho}&=\Phi_2^{(0)}(s,m_1,m_2)\left[ A_3 p^{\mu}p^{\nu}p^{\rho}+B_3 p^2 p^{\mu}g^{\nu\rho}+C_3 p^2\left(g^{\mu\nu}p^{\rho}+g^{\mu\rho}p^{\nu}\right)\right],
\end{align}
where $A_i$, $B_i$ and $C_i$ are functions of $s$, $m_1$ and $m_2$. Their expressions are
\begin{align}
A_1&=\frac{m_1^2-m_2^2+s}{2 s},\\
A_2&=-\frac{ s \left(-m_1^2-m_2^2+s\right)-2\left(m_1^2-m_2^2+s\right)\left(-m_1^2+m_2^2+s\right)}{6 s^2},\\
B_2&=-\frac{\left(m_1^2-m_2^2+s\right)\left(-m_1^2+m_2^2+s\right)-2 s\left(-m_1^2-m_2^2+s\right)}{12 s^2},\\
A_3&=-\frac{1}{3 s^3}\left[\frac{1}{2} m_1^2 s\left(m_1^2-m_2^2+s\right)-\frac{3}{4}\left(m_1^2-m_2^2+s\right)\left(-m_1^2+m_2^2+s\right)^2\right.\nonumber\\
&\left.+\frac{1}{2} s\left(-m_1^2-m_2^2+s\right)\left(-m_1^2+m_2^2+s\right)\right],\\
B_3&=\frac{\left(m_1^2-m_2^2+s\right) \left(4 m_1^2s-\left(-m_1^2+m_2^2+s\right)^2\right)}{24 s^3},\\
C_3&=-\frac{1}{3 s^3}\left[\frac{1}{8}\left(m_1^2-m_2^2+s\right)\left(-m_1^2+m_2^2+s\right)^2-\frac{1}{4} s\left(-m_1^2-m_2^2+s\right)\left(-m_1^2+m_2^2+s\right)\right].
\end{align}

\section{Expression of the ${\cal M}, {\cal N}$ and ${\cal L}$ functions\label{sec:LCDAofBD}}
In this appendix, we list all the ${\cal M}, {\cal N}$ and ${\cal L}$ functions appearing in the imaginary part of the correlation function for $\langle \pi^- (S, V, T) \left |{\cal O}_1\right |B^-\rangle$. For the case of scalar resonance, we have
\begin{align}
&\int d\Phi_2(p_S^2,k_2,k_3)\ {\cal M}^{S}(p_{\pi},k_2,k_3)\nonumber\\
=&-\frac{1}{m_B}\left[8 \left(2 A_2 \left(-m_B^3+m_B^2 \omega +m_B
   m_{\pi}^2+p_{\pi}^2 \omega \right)+(4 B_2-1) p_{\pi}^2 \omega \right)\right],\\
   &\int d\Phi_2(p_S^2,k_2,k_3)\ {\cal N}^{S}(p_{\pi},k_2,k_3)\nonumber\\
=&\frac{1}{m_B}\left[32 \left(2 A_2 \left(m_B^2+p_{\pi}^2\right)+(4
   B_2-1) p_{\pi}^2\right)\right],\\
   &\int d\Phi_2(p_S^2,k_2,k_3)\ {\cal L}^{S}(p_{\pi},k_2,k_3)\nonumber\\
=&-\frac{1}{m_B^2}\left[16 \left(2 A_2 \left(m_B^2+p_{\pi}^2\right)+(4
   B_2-1) p_{\pi}^2\right) \left(m_B^2 \omega -m_B
   \left(\text{pVs}+\omega ^2\right)+\omega 
   \left(\text{pVs}-m_{\pi}^2\right)\right)\right].
\end{align}
For the case of vector resonance, we have
\begin{align}
&\int d\Phi_2(p_V^2,k_2,k_3)\ {\cal M}^{V(1)}_{(a)}(p_{\pi},k_2,k_3)\nonumber\\
=&-\frac{8}{m_B^2}\left[A_2\left(m_B^{2}-m_{\pi}^{2}+p_V^2\right)\left(2m_B^{3}-2m_B^{2}\omega+m_B\left(p_V^2-2m_{\pi}^{2}\right)+2\omega\left(m_{\pi}^{2}-p_V^2\right)\right)\right.\nonumber\\
&\left.+(4B_2-1)m_B^{2}p_V^2(m_B-\omega)\right],\\
\nonumber\\
&\int d\Phi_2(p_V^2,k_2,k_3)\ {\cal M}^{V(2)}_{(a)}(p_{\pi},k_2,k_3)\nonumber\\
=&-\frac{1}{m_B^2}\left[4p_V^2\left(m_B^{3}(4A_2+4B_2-1)-2m_B^{2}\omega(2A_2+4B_2-1)\right.\right.\nonumber\\
&\left.\left.+m_B\left(A_2\left(2p_V^2-4m_{\pi}^{2}\right)+(4B_2-1)(p_V^2-p_{\pi}^2)\right)+2\omega\left(2A_2\left(m_{\pi}^{2}-p_V^2\right)+(4B_2-1)p_{\pi}^2\right)\right)\right],\\
\nonumber\\
&\int d\Phi_2(p_V^2,k_2,k_3)\ {\cal N}^{V(1)}_{(a)}(p_{\pi},k_2,k_3)= -\frac{32A_1(m_B-\omega)\left(m_B^{2}-m_{\pi}^{2}+p_V^2\right)}{m_B},\\
\nonumber\\
&\int d\Phi_2(p_V^2,k_2,k_3)\ {\cal N}^{V(2)}_{(a)}(p_{\pi},k_2,k_3)=\frac{32A_1p_V^2(\omega-m_B)}{m_B},\\
\nonumber\\
&\int d\Phi_2(p_V^2,k_2,k_3)_{\Omega}\ {\cal L}^{V(1)}_{(a)}(p_{\pi},k_2,k_3)\nonumber\\
=&-\frac{1}{m_B}\left[16\left(m_B^{3}(p_V^2(A_1-2B_3+2C_3-1)-A_1\Omega_2+\Omega_2)\right.\right.\nonumber\\
&\left.\left.+m_B^{2}\omega((A_1-1)\Omega_2-p_V^2(A_1-2B_3+2C_3-1))\right.\right.\nonumber\\
&\left.\left.-m_B\left(m_{\pi}^{2}(p_V^2(A_1-2B_3+2C_3-1)-A_1\Omega_2+\Omega_2)+A_1p_V^2\Omega_2\right)\right.\right.\nonumber\\
&\left.\left.+\omega\left(m_{\pi}^{2}-p_V^2\right)(p_V^2(A_1-2B_3+2C_3-1)-A_1\Omega_2+\Omega_2)\right)\right],\\
\nonumber\\
&\int d\Phi_2(p_V^2,k_2,k_3)_{\Omega}\ {\cal L}^{V(2)}_{(a)}(p_{\pi},k_2,k_3)=16 A_1 p_V^2\Omega\left(1-\frac{\omega}{m_B}\right),\\
&\int d\Phi_2(p_V^2,k_2,k_3)\ {\cal M}^{V(1)}_{(b)}(p_{\pi},k_2,k_3)\nonumber\\
=&\ 8 \left(2 A_2 \left(m_B^3-m_B^2 \omega -m_B m_{\pi}^2-p_{\pi}^2\omega \right)+(1-4 B_2)p_{\pi}^2 \omega\right) \omega,\\
\nonumber\\
&\int d\Phi_2(p_V^2,k_2,k_3)\ {\cal M}^{V(2)}_{(b)}(p_{\pi},k_2,k_3)\nonumber\\
=&\frac{4}{m_B^2} \left(2 A_2\left(m_B^5-2 m_B^4\omega -m_B^3\left(m_{\pi}^2+p_{\pi}^2\right)+2 m_B^2 p_V^2 \omega\right.\right.\nonumber\\
&\left.\left. +m_B \left(m_{\pi}^2 (p_{\pi}^2+p_V^2)+p_V^2(p_{\pi}^2-p_V^2)\right)+2 p_{\pi}^4 \omega\right)+(4 B_2-1)p_{\pi}^2 \left(-2 m_B^2 \omega +m_B p_V^2+2 p_{\pi}^2\omega\right)\right),\\
\nonumber\\
&\int d\Phi_2(p_V^2,k_2,k_3)\ {\cal N}^{V(1)}_{(b)}(p_{\pi},k_2,k_3)=-32 A_2 m_B^2-32 A_2 p_{\pi}^2-64 B_2
   p_{\pi}^2+16 p_{\pi}^2,\\
&\int d\Phi_2(p_V^2,k_2,k_3)\ {\cal N}^{V(2)}_{(b)}(p_{\pi},k_2,k_3)=-32 A_2 p_V^2,\\
&\int d\Phi_2(p_V^2,k_2,k_3)_{\Omega}\ {\cal L}^{V(1)}_{(b)}(p_{\pi},k_2,k_3)\nonumber\\
=&-16 \left(2 A_2\left(m_B^4-2 m_B^3\omega +m_B^2 \left(\omega^2-2 m_{\pi}^2\right)+2 m_B m_{\pi}^2 \omega+m_{\pi}^4-m_{\pi}^2 p_V^2+p_{\pi}^2 \omega^2-p_{\pi}^2 p_V^2\right)\right.\nonumber\\
&\left.-(4 B_2-1) p_{\pi}^2 \left(p_V^2-\omega^2\right)\right),\\
\nonumber\\
&\int d\Phi_2(p_V^2,k_2,k_3)_{\Omega}\ {\cal L}^{V(2)}_{(b)}(p_{\pi},k_2,k_3)\nonumber\\
=&\ \frac{1}{m_B^2}\left[16 \left(2 A_2 \left(m_B^5\omega -m_B^4 \left(p_V^2+\omega^2\right)-m_B^3 \omega \left(m_{\pi}^2+p_{\pi}^2\right)+m_B^2 p_V^2\left(m_{\pi}^2+p_V^2+\omega ^2\right)\right.\right.\right.\nonumber\\
&\left.\left.\left.+m_B \omega \left(m_{\pi}^2 (p_{\pi}^2+p_V^2)+p_V^2(p_{\pi}^2-p_V^2)\right)+p_{\pi}^4 \omega ^2\right)+(4B_2-1) p_{\pi}^2\omega  \left(m_B^2(-\omega )+m_Bp_V^2+p_{\pi}^2 \omega \right)\right)\right].
\end{align}
For the case of tensor resonance, we have
\begin{align}
&\int d\Phi_2(p_V^2,k_2,k_3)\ {\cal M}^{T(1)}_{(a)}(p_{\pi},k_2,k_3)\nonumber\\
=&\frac{1}{m_B^2}\left[4 \left(m_B^2  p_T^2 \left(m_B^3 (-2
   A_1+4 B_2-12 B_3-20 C_3+1)\right.\right.\right.\nonumber\\
   &\left.\left.\left.+m_B^2 \omega 
   (2 A_1-4 B_2+12 B_3+20 C_3-1)\right.\right.\right.\nonumber\\
   &\left.\left.\left.+m_B
   \left(m_{\pi}^2 (2 A_1-4 B_2+12 B_3+20
   C_3-1)+ p_T^2 (-2 A_1+4 B_2-8 B_3-16
   C_3+1)\right)\right.\right.\right.\nonumber\\
   &\left.\left.\left.-\omega  \left(m_{\pi}^2- p_T^2\right) (2
   A_1-4 B_2+12 B_3+20 C_3-1)\right)\right.\right.\nonumber\\
   &\left.\left.+A_2
   \left(m_B^2-m_{\pi}^2+ p_T^2\right)^2 \left(2
   m_B^3-2 m_B^2 \omega +m_B \left( p_T^2-2
   m_{\pi}^2\right)+2 \omega 
   \left(m_{\pi}^2- p_T^2\right)\right)\right.\right.\nonumber\\
   &\left.\left.-2 A_3
   \left(m_B^2-m_{\pi}^2+ p_T^2\right)^2 \left(2
   m_B^3-2 m_B^2 \omega +m_B \left( p_T^2-2
   m_{\pi}^2\right)+2 \omega 
   \left(m_{\pi}^2- p_T^2\right)\right)\right)\right],\\
   &\int d\Phi_2(p_V^2,k_2,k_3)\ {\cal M}^{T(2)}_{(a)}(p_{\pi},k_2,k_3)\nonumber\\
=&\frac{1}{m_B^2}\left[p_T^2 \left(m_B^5 (-2 A_1+8 A_2-16 A_3+8
   B_2-16 B_3-32 C_3+1)\right.\right.\nonumber\\
   &\left.\left.+2 m_B^4 \omega  (2
   A_1-4 A_2+8 A_3-8 B_2+8 B_3+24
   C_3-1)\right.\right.\nonumber\\
   &\left.\left.+m_B^3 \left(m_{\pi}^2 (2 A_1-16
   A_2+32 A_3-8 B_2+16 B_3+32 C_3-1)\right.\right.\right.\nonumber\\
   &\left.\left.\left.-4
    p_T^2 (2 A_1-3 A_2+6 A_3-4 B_2+6
   B_3+14 C_3-1)+p_{\pi}^2 (2 A_1-8 B_2+16
   B_3+32 C_3-1)\right)\right.\right.\nonumber\\
   &\left.\left.-2 m_B^2 \omega 
   \left(m_{\pi}^2 (2 A_1-8 A_2+16 A_3-8 B_2+8
   B_3+24 C_3-1)\right.\right.\right.\nonumber\\
   &\left.\left.\left.-2  p_T^2 (2 A_1-4 A_2+8
   A_3-3 B_2+9 B_3+15 C_3-1)+p_{\pi}^2 (2
   A_1-8 B_2+8 B_3+24 C_3-1)\right)\right.\right.\nonumber\\
   &\left.\left.+m_B
   \left(-2 A_1 m_{\pi}^2 p_{\pi}^2+2 A_1
   m_{\pi}^2  p_T^2-2 A_1   p_T^4+2 A_1
   p_{\pi}^2  p_T^2+4 A_2 \left(2 m_{\pi}^4-3
   m_{\pi}^2  p_T^2+  p_T^4\right)\right.\right.\right.\nonumber\\
   &\left.\left.\left.-8 A_3 \left(2
   m_{\pi}^4-3 m_{\pi}^2  p_T^2+  p_T^4\right)+8 B_2
   m_{\pi}^2 p_{\pi}^2-8 B_2 m_{\pi}^2  p_T^2+8
   B_2   p_T^4-8 B_2 p_{\pi}^2  p_T^2\right.\right.\right.\nonumber\\
   &\left.\left.\left.-16
   B_3 m_{\pi}^2 p_{\pi}^2+8 B_3 p_{\pi}^2
    p_T^2-32 C_3 m_{\pi}^2 p_{\pi}^2+16 C_3
   m_{\pi}^2  p_T^2-16 C_3   p_T^4\right.\right.\right.\nonumber\\
   &\left.\left.\left.+24 C_3
   p_{\pi}^2  p_T^2+m_{\pi}^2 p_{\pi}^2-m_{\pi}^2
    p_T^2+  p_T^4-p_{\pi}^2  p_T^2\right)\right.\right.\nonumber\\
   &\left.\left.+2 \omega 
   \left(m_{\pi}^2- p_T^2\right) \left(p_{\pi}^2 (2
   A_1-8 B_2+8 B_3+24 C_3-1)\right.\right.\right.\nonumber\\
   &\left.\left.\left.-4 A_2
   \left(m_{\pi}^2- p_T^2\right)+8 A_3
   \left(m_{\pi}^2- p_T^2\right)\right)\right)\right],\\
   &\int d\Phi_2(p_V^2,k_2,k_3)\ {\cal M}^{T(3)}_{(a)}(p_{\pi},k_2,k_3)\nonumber\\
=&-\frac{1}{m_B^2}\left[2   p_T^4 \left(m_B^3 (2 A_1-4 A_2+8
   A_3-2 B_2+6 B_3+10 C_3-1)\right.\right.\nonumber\\
   &\left.\left.-2 m_B^2 \omega 
   (2 A_1-2 A_2+4 A_3-2 B_2+6 B_3+10
   C_3-1)\right.\right.\nonumber\\
   &\left.\left.+m_B \left(-2 A_1 p_{\pi}^2+2 A_1
    p_T^2+4 A_2 m_{\pi}^2-2 A_2  p_T^2-8 A_3
   m_{\pi}^2+4 A_3  p_T^2+2 B_2 p_{\pi}^2\right.\right.\right.\nonumber\\
   &\left.\left.\left.-6
   B_2  p_T^2-6 B_3 p_{\pi}^2+2 B_3
    p_T^2-10 C_3 p_{\pi}^2+14 C_3
    p_T^2+p_{\pi}^2- p_T^2\right)\right.\right.\nonumber\\
   &\left.\left.+2 \omega 
   \left(p_{\pi}^2 (2 A_1-2 B_2+6 B_3+10
   C_3-1)-2 A_2 \left(m_{\pi}^2- p_T^2\right)+4
   A_3
   \left(m_{\pi}^2- p_T^2\right)\right)\right)\right],\\
    &\int d\Phi_2(p_V^2,k_2,k_3)\ {\cal N}^{T(1)}_{(a)}(p_{\pi},k_2,k_3)\nonumber\\
=&-\frac{1}{m_B}\left[16 (m_B-\omega ) \left(A_1
   \left(m_B^2-m_{\pi}^2+p_T^2\right)^2-2 \left(A_2
   \left(m_B^2-m_{\pi}^2+p_T^2\right)^2+4 B_2
   m_B^2 p_T^2\right)\right)\right],\\
   &\int d\Phi_2(p_V^2,k_2,k_3)\ {\cal N}^{T(2)}_{(a)}(p_{\pi},k_2,k_3)\nonumber\\
=&\frac{1}{m_B}\left[16 p_T^2
   (m_B-\omega ) \left(A_1
   \left(m_B^2-m_{\pi}^2+p_T^2\right)-2 \left(A_2
   \left(m_B^2-m_{\pi}^2+p_T^2\right)+2 B_2
   \left(m_B^2-p_{\pi}^2\right)\right)\right)\right],\\
   &\int d\Phi_2(p_V^2,k_2,k_3)\ {\cal N}^{T(3)}_{(a)}(p_{\pi},k_2,k_3)\nonumber\\
=&\frac{1}{m_B}\left[16 p_T^4 (A_1-2
   A_2) (m_B-\omega )\right],\\
    &\int d\Phi_2(p_V^2,k_2,k_3)_{\Omega}\ {\cal L}^{T(1)}_{(a)}(p_{\pi},k_2,k_3)\nonumber\\
=&\frac{1}{m_B}\left[8 p_T^2 \left(m_B^2-m_{\pi}^2+p_T^2\right)
   \left(-m_B^3+m_B^2 \omega +m_B m_{\pi}^2+\omega 
   \left(p_T^2-m_{\pi}^2\right)\right)\right.\nonumber\\
   &\left. \times(A_1+2 A_2+2
   B_2+4 B_3+4 C_3-1)\right],\\
   &\int d\Phi_2(p_V^2,k_2,k_3)_{\Omega}\ {\cal L}^{T(2)}_{(a)}(p_{\pi},k_2,k_3)\nonumber\\
=&\frac{1}{m_B}\left[4 p_T^4
   \left(m_B^3 (-(A_1+2 A_2-4 B_2-2 B_3+10
   C_3-1))\right.\right.\nonumber\\
   &\left.\left.+m_B \left(m_{\pi}^2 (A_1+2 A_2-4
   B_2-2 B_3+10 C_3-1)+4 p_T^2
   (B_2+B_3-C_3)\right)\right.\right.\nonumber\\
   &\left.\left.+m_B^2 \omega  (A_1+2
   A_2+2 B_3+6 C_3-1)-\omega 
   \left(m_{\pi}^2-p_T^2\right) (A_1+2 A_2+2
   B_3+6 C_3-1)\right)\right],\\
   &\int d\Phi_2(p_V^2,k_2,k_3)_{\Omega}\ {\cal L}^{T(3)}_{(a)}(p_{\pi},k_2,k_3)=0,\\
   &\int d\Phi_2(p_V^2,k_2,k_3)\ {\cal M}^{T(1)}_{(b)}(p_{\pi},k_2,k_3)\nonumber\\
=&4 \left(m_B^2-2 m_B \omega
   -m_{\pi}^2+p_T^2\right) \left(2 A_2
   \left(m_B^3-m_B^2 \omega -m_B
   m_{\pi}^2-p_{\pi}^2 \omega \right)+(1-4 B_2) p_{\pi}^2 \omega \right),\\
   &\int d\Phi_2(p_V^2,k_2,k_3)\ {\cal M}^{T(2)}_{(b)}(p_{\pi},k_2,k_3)\nonumber\\
=&\frac{1}{m_B^2}\left[2 A_2 \left(m_B^7-6 m_B^6
   \omega -m_B^5 \left(2 m_{\pi}^2+p_{\pi}^2-3 p_T^2-6
   \omega ^2\right)\right.\right.\nonumber\\
   &\left.\left.+2 m_B^4 \omega  \left(3 m_{\pi}^2+2
   p_{\pi}^2-p_T^2\right)+m_B^3 \left(m_{\pi}^4-2
   m_{\pi}^2 (p_T^2-p_{\pi}^2)-p_T^2 \left(p_T^2+4
   \omega ^2\right)\right)\right.\right.\nonumber\\
   &\left.\left.+2 m_B^2 \omega  \left(-2 m_{\pi}^2
   (p_{\pi}^2+p_T^2)+p_{\pi}^4+2 p_T^4-2
   p_{\pi}^2 p_T^2\right)\right.\right.\nonumber\\
   &\left.\left.-m_B \left(m_{\pi}^4
   (p_{\pi}^2+p_T^2)-2 m_{\pi}^2 p_T^4+6 p_{\pi}^4 \omega ^2+p_T^6-p_{\pi}^2 p_T^4\right)+2
   p_{\pi}^4 \omega 
   \left(p_T^2-m_{\pi}^2\right)\right)\right.\nonumber\\
   &\left.-(4 B_2-1) p_{\pi}^2 \left(2 m_B^4 \omega -m_B^3 \left(p_T^2+6 \omega
   ^2\right)-2 m_B^2 \omega  \left(m_{\pi}^2+p_{\pi}^2-3
   p_T^2\right)\right.\right.\nonumber\\
   &\left.\left.+m_B \left(m_{\pi}^2 p_T^2+6 p_{\pi}^2 \omega ^2-p_T^4\right)+2 p_{\pi}^2 \omega 
   \left(m_{\pi}^2-p_T^2\right)\right)\right],\\
   &\int d\Phi_2(p_V^2,k_2,k_3)\ {\cal M}^{T(3)}_{(b)}(p_{\pi},k_2,k_3)\nonumber\\
=&\frac{1}{m_B^3}\left[2
   \left(m_B^2 \omega -m_B p_T^2-p_{\pi}^2 \omega
   \right) \left(2 A_2 \left(m_B^5-2 m_B^4 \omega
   -m_B^3 \left(m_{\pi}^2+p_{\pi}^2\right)\right.\right.\right.\nonumber\\
   &\left.\left.\left.+2 m_B^2
   p_T^2 \omega +m_B \left(m_{\pi}^2 (p_{\pi}^2+p_T^2)+p_T^2 (p_{\pi}^2-p_T^2)\right)\right.\right.\right.\nonumber\\
   &\left.\left.\left.+2
   p_{\pi}^4 \omega \right)+(4 B_2-1) p_{\pi}^2
   \left(-2 m_B^2 \omega +m_B p_T^2+2 p_{\pi}^2
   \omega \right)\right)\right],\\
   &\int d\Phi_2(p_V^2,k_2,k_3)\ {\cal N}^{T(1)}_{(b)}(p_{\pi},k_2,k_3)\nonumber\\
=&-8 \left(2 A_2 \left(m_B^2+p_{\pi}^2\right)+(4
   B_2-1) p_{\pi}^2\right) \left(m_B^2-2 m_B
   \omega -m_{\pi}^2+p_T^2\right),\\
   &\int d\Phi_2(p_V^2,k_2,k_3)\ {\cal N}^{T(2)}_{(b)}(p_{\pi},k_2,k_3)\nonumber\\
=&\frac{1}{m_B}\left[4 \left(2 A_2
   \left(m_B^4 \omega -2 m_B^3 p_T^2+2 m_B^2
   p_T^2 \omega -m_B p_T^2 \left(-m_{\pi}^2+p_{\pi}^2+p_T^2\right)-p_{\pi}^4 \omega \right)\right.\right.\nonumber\\
   &\left.\left.-(4 B_2-1)
   p_{\pi}^2 \left(-\omega m_B^2+m_B
   p_T^2+p_{\pi}^2 \omega \right)\right)\right],\\
   &\int d\Phi_2(p_V^2,k_2,k_3)\ {\cal N}^{T(3)}_{(b)}(p_{\pi},k_2,k_3)\nonumber\\
=&-\frac{1}{m_B}\left[16
   A_2 p_T^2 \left(-\omega m_B^2 +m_B
   p_T^2+p_{\pi}^2 \omega \right)\right],\\
   &\int d\Phi_2(p_V^2,k_2,k_3)_{\Omega}\ {\cal L}^{T(1)}_{(b)}(p_{\pi},k_2,k_3)\nonumber\\
=&4 \left(m_B^2-2 m_B \omega
   -m_{\pi}^2+p_T^2\right) \left(2 A_2 \left(m_B^4-2
   m_B^3 \omega +m_B^2 \left(\omega ^2-2 m_{\pi}^2\right)\right.\right.\nonumber\\
   &\left.\left.+2
   m_B m_{\pi}^2 \omega +m_{\pi}^4-m_{\pi}^2
   p_T^2+p_{\pi}^2 \omega ^2-p_{\pi}^2
   p_T^2\right)-(4 B_2-1) p_{\pi}^2
   \left(p_T^2-\omega ^2\right)\right),\\
    &\int d\Phi_2(p_V^2,k_2,k_3)_{\Omega}\ {\cal L}^{T(2)}_{(b)}(p_{\pi},k_2,k_3)\nonumber\\
=&\frac{1}{m_B^2}\left[2 \left(2 A_2
   \left(2 m_B^7 \omega -m_B^6 \left(2 p_T^2+5 \omega
   ^2\right)+m_B^5 \omega  \left(-4 m_{\pi}^2-2 p_{\pi}^2+5
   p_T^2+3 \omega ^2\right)\right.\right.\right.\nonumber\\
   &\left.\left.\left.+m_B^4 \left(m_{\pi}^2 \left(4
   p_T^2+5 \omega ^2\right)-\omega ^2 (p_T^2-4 p_{\pi}^2)\right)\right.\right.\right.\nonumber\\
   &\left.\left.\left.-m_B^3 \omega  \left(-2 m_{\pi}^4+m_{\pi}^2 (5
   p_T^2-4 p_{\pi}^2)+p_T^2 \left(p_{\pi}^2+3
   p_T^2+2 \omega ^2\right)\right)\right.\right.\right.\nonumber\\
   &\left.\left.\left.+m_B^2 \left(-2 m_{\pi}^4
   p_T^2+m_{\pi}^2 \left(-4 p_{\pi}^2 \omega
   ^2+p_T^4-3 p_T^2 \omega ^2\right)+p_{\pi}^4 \omega
   ^2+p_T^6+p_T^4 \left(p_{\pi}^2+3 \omega ^2\right)-3
   p_{\pi}^2 p_T^2 \omega ^2\right)\right.\right.\right.\nonumber\\
   &\left.\left.\left.+m_B \omega 
   \left(m_{\pi}^4 (-(2 p_{\pi}^2+p_T^2))+m_{\pi}^2
   p_T^2 (p_{\pi}^2+2 p_T^2)-3 p_{\pi}^4 \omega
   ^2-p_T^6+p_{\pi}^2 p_T^4+p_{\pi}^4
   p_T^2\right)\right.\right.\right.\nonumber\\
   &\left.\left.\left.+p_{\pi}^4 \omega ^2
   \left(p_T^2-m_{\pi}^2\right)\right)\right.\right.\nonumber\\
   &\left.\left.+(4 B_2-1) p_{\pi}^2 \left(m_B^2 \omega +m_B \left(p_T^2-3 \omega
   ^2\right)+\omega  \left(p_T^2-m_{\pi}^2\right)\right)
   \left(m_B^2 (-\omega )+m_B p_T^2+p_{\pi}^2
   \omega \right)\right)\right],\\
   &\int d\Phi_2(p_V^2,k_2,k_3)_{\Omega}\ {\cal L}^{T(3)}_{(b)}(p_{\pi},k_2,k_3)\nonumber\\
=&\frac{1}{m_B^3}\left[4 \left(m_B^2 \omega
   -m_B p_T^2-p_{\pi}^2 \omega \right) \left(2 A_2
   \left(m_B^5 \omega -m_B^4 \left(p_T^2+\omega
   ^2\right)\right.\right.\right.\nonumber\\
   &\left.\left.\left.-m_B^3 \omega  \left(m_{\pi}^2+p_{\pi}^2\right)+m_B^2 p_T^2 \left(m_{\pi}^2+p_T^2+\omega
   ^2\right)+m_B \omega  \left(m_{\pi}^2 (p_{\pi}^2+p_T^2)+p_T^2 (p_{\pi}^2-p_T^2)\right)+p_{\pi}^4 \omega ^2\right)\right.\right.\nonumber\\
   &\left.\left.+(4
   B_2-1) p_{\pi}^2 \omega  \left(-m_B^2 \omega
   +m_B p_T^2+p_{\pi}^2 \omega
   \right)\right)\right].
\end{align}

\end{appendix}

\end{document}